\documentclass[12pt]{article}

\usepackage[margin=1in]{geometry}
\usepackage{amsmath,amssymb,bm,bbm}
\usepackage{amsthm}
\usepackage{hyperref}
\usepackage{footnote}
\usepackage{natbib}
\usepackage{comment}
\usepackage{kotex}
\usepackage[dvipsnames]{xcolor}
\usepackage{dsfont}
\usepackage{mathtools}
\usepackage{tikz}
\usepackage{color}
\usepackage{algorithm}
\usepackage{algpseudocode}
\usepackage{dsfont}
\usepackage{graphicx,subcaption}
\usepackage{booktabs}
\usepackage{appendix}
\usepackage{caption}
\usepackage{lineno}
\usepackage{bbold}
\usepackage{fancybox}
\usepackage{enumitem}
\usepackage{tikz}
\usetikzlibrary{arrows.meta, positioning}

\newtheorem{theorem}{Theorem}
\newtheorem{lemma}{Lemma}
\newtheorem{corollary}{Corollary}

\newtheorem{assumption}{Assumption}
\newtheorem*{remark}{Remark}
 
\DeclareMathOperator*{\argmax}{arg\,max}

\renewcommand{\k}{\tilde{k}}
\renewcommand{\l}{\tilde{l}}

\renewcommand{\P}{\mathbb{P}}

\newcommand{\ZZ}{\mathcal{Z}}
\renewcommand{\SS}{\hat{\mathcal{S}}}
\renewcommand{\P}{\mathrm{P}}

\newcommand{\ind}{\perp\!\!\!\!\perp} 

\DeclarePairedDelimiter\floor{\lfloor}{\rfloor}

\newcommand{\papertitle}{Semi-Supervised Learning of Noisy Mixture of
  Experts Models}

\def\spacingset#1{\renewcommand{\baselinestretch}%
{#1}\small\normalsize} \spacingset{1}

\begin{document}

  \title{\bf \papertitle}
  \author{Oh-Ran Kwon, Gourab Mukherjee, and Jacob Bien
    \hspace{.2cm}\\
    Data Sciences and Operations, University of Southern California}
  \maketitle
  
\begin{abstract}

  The mixture of experts (MoE) model is a versatile framework for predictive modeling that has gained renewed interest in the age of large language models.  A collection of predictive ``experts'' is learned along with a ``gating function'' that controls how much influence each expert is given when a prediction is made.  This structure allows relatively simple models to excel in complex, heterogeneous data settings.

  In many contemporary settings, unlabeled data are widely available while labeled data are difficult to obtain.  Semi-supervised learning methods seek to leverage the unlabeled data. 
  We propose a novel method for semi-supervised learning of MoE models.  We start from a semi-supervised MoE model that was developed by oceanographers that makes the strong assumption that the latent clustering structure in unlabeled data maps directly to the influence that the gating function should give each expert in the supervised task.  We relax this assumption, imagining a noisy connection between the two, and propose an algorithm based on least trimmed squares, which succeeds even in the presence of misaligned data. Our theoretical analysis characterizes the conditions under which our approach yields estimators with a near-parametric rate of convergence. Simulated and real data examples demonstrate the method's efficacy. 
  \end{abstract}

  \noindent%
\vfill

\section{Introduction}

Many contemporary prediction problems have two prominent features: (i) data are heterogenous, influenced by phenomena that are difficult to describe in a simple functional form and (ii) an abundance of data is available that is loosely related but not precisely tailored to the prediction task one wishes to solve.  In this work, we propose a method to tackle the challenge in (i) by leveraging the opportunity in (ii).

When modeling heterogenous data and complex phenomena, it can be tempting to assume one needs extremely sophisticated, all-encompassing, generalist models.  However, parsimony can still sometimes be obtained in such settings using a small collection of simple specialist models.  The challenge becomes learning which specialist model(s) to rely on in which settings.  The mixture of experts (MoE) model \citep{jacobs1991adaptive,jordan1994hierarchical} is a framework for simultaneously learning these specialist models (``experts'') and the ``gate function'' that is responsible for using the input value to determine how much weight each expert is given. 
This framework has been used extensively across many application areas over the last few decades \citep{yuksel2012twenty}. In recent years, the MoE model has garnered fresh attention as it has achieved notable success within deep learning architectures. 
The extension of the MoE model to deep learning architectures was introduced by \cite{eigen2013learning}, who conceptualized an MoE model as a layer and proposed stacking multiple mixture of experts layers. Since then, various MoE layer variants have been proposed and successfully applied to tasks in large language models. For example, \cite{shazeer2017outrageously} introduced the sparsely-gated layer, which routes an input to a small subset of experts, while \citet{fedus2022switch} proposed the switch layer, which routes an input to a single expert. Additionally, \citealt{pham2024competesmoe} introduced CompeteSMoE, which routes an input only to a small number of experts providing the highest outputs. 
In the statistics literature, there has been some effort directed toward high-dimensional settings. For example, there is an approach that utilizes inverse regression methods, which exchange the roles of input and response, providing a parsimonious representation of the parameters \citep{deleforge2015high,nguyen2022non}. Additionally, penalized maximum likelihood estimation is considered \citep{khalili2010new,hyun2023modeling}. 
Recently, \cite{javanmard2022prediction} introduced an approach to construct prediction sets for the high-dimensional MoE model using a debiasing technique to account for the bias induced by penalization.

In many domains of application, labeled data $(X,Y)\in\mathbb R^p \times \mathbb R$ are costly and rare, while unlabeled data $X$ are abundant and easily accessible. For example, oceanographers must make resource-intensive voyages out to sea to measure important quantities such as chlorophyll concentration; such data that they can collect is much less abundant than satellite measurements, which are collected nearly daily and provide global coverage of the ocean \citep{lee2006remote}.  Likewise, in chemometrics, spectroscopy data are quick and cheap to obtain, while measuring chemical properties require time-intensive, traditional laboratory methods \citep{barra2021soil}. In natural language processing, relatively few labeled data are available for many tasks (e.g., question-answer pairs), but vast quantities of raw text can be easily obtained from the Internet \citep{tsvetkov2017lowresource}. Semi-supervised learning \citep{zhu2005semi,chapelle2006semi} sets out to exploit unlabeled data when it is abundantly available with the goal of improving performance on supervised tasks.  A key breakthrough for current large language models came from the use of semi-supervised learning \citep{bengio2006greedy,JMLR:v11:erhan10a}, whose importance is highlighted in the phrase ``pre-trained'' of ``generative pre-trained transformer'' (GPT, \citealt{radford2018improving}).

Considering the power of the MoE framework for handling heterogenous data and  the great potential of vast unlabeled data sets, we develop here a new semi-supervised learning approach for training MoE models. Prior work on semi-supervised training of MoE models appears limited, especially in the regression setting.
Since semi-supervised learning typically focuses on classification tasks, we are aware of a few studies in the classification setting. Early studies employed the expectation-maximization algorithm to maximize the likelihood-based function given both labeled and unlabeled data \citep{miller1996mixture,karakoulas2004semi,miller2009semisupervised}. Recent studies adopted machine learning techniques with more complex models for the experts \citep{kiyono2019mixture,kizaric2022classifying}.

In the oceanography literature, the state-of-the-art method for predicting chlorophyll concentration from multispectral satellite data can be interpreted as a semi-supervised mixture of experts model (although the authors use none of these terms).  In particular, \cite{moore2001fuzzy} proposed a two-step algorithm, which has been adopted by \cite{jackson2017improved,hieronymi2017olci}. Abstracting away some of the detailed choices they made, we describe the two steps as (i) estimating the gate function using unlabeled data only and (ii)  estimating the experts using labeled data and some information from the gate function. Details of their algorithm are summarized in Section \ref{subsec:lit_method}. Their approach is based on the strong assumption that the latent membership of an observation $X$ to an (unsupervised) cluster provides perfect information about which expert to use in the supervised task. In other words, if we illustrate their approach under Gaussian model assumptions, it can be thought of arising from what we will refer to as the {\em semi-supervised MoE model}: 
\begin{equation}\label{eq:ssmoe}
    \begin{aligned}
        &{\color{black}Y = \beta_{k0} + \beta_k^T X + \epsilon_k, \text{ for } Z=k,}   \\
        \text{and~}&X|(Z=k) \sim N(\mu_{k}, \Sigma_{k}), \text{ for } k = 1,\ldots, K,  
    \end{aligned}
\end{equation}
where {\color{black}$\epsilon_k\in\mathbb R$ is the error term} and each expert is the distribution of $Y|X,Z$. We can view this model as establishing clustering structures on $Y|X$ and on $X$, which are both based on the same latent variable $Z$. In this case, understanding the distribution of $(Z,X)$ from abundant unlabeled data can help identify the clustering structure in $Y|X$, which can provide useful information in supervised problems. While this assumption may hold true in their specific oceanography application, it restricts the applicability of their approach to other problems.
 
In this paper, we introduce the {\em noisy semi-supervised MoE model} (which we call ``noisy MoE'' for short) by weakening their assumption: the clustering structure of $X$ is similar but not necessarily identical to the clustering structure of $Y|X$. Specifically, we define the noisy MoE with the following model:
\begin{equation*}\label{eq:noisymoe}
    \begin{aligned}
        &{\color{black}Y = \beta_{k0} + \beta_k^T X + \epsilon_k, \text{~for~} Z=k,}   \\
        \text{and~}&X|(\tilde Z=\k) \sim N(\mu_{\k}, \Sigma_{\k}), \text{ for } k,\k = 1,\ldots, K,
    \end{aligned}
\end{equation*}
and $\mathrm P(Z \neq \tilde Z)$ is thought to be a small value. 
We use the word ``noisy" in the name of this model because we can think of $Z$, which captures the unsupervised clustering structure, as being a noisy version of $\tilde Z$, which is the latent variable we care about for the supervised part of the problem. 
In the noisy MoE, the transition probabilities for going from $\tilde{Z}$ to $Z$ is a key component that needs to be modeled and estimated. 
A schematic of the 
model and the data considered in this paper is presented in Figure~\ref{fig:relation}. 

We develop a semi-supervised approach for the noisy MoE. As in the semi-supervised approach for MoE, 
we assign each labeled data point to the clusters defined by $\tilde Z$ based on the estimated conditional distribution of $\tilde Z|X$ using a large amount of unlabeled data.
In the clustering structure of $X$
defined by $Z$, however, some data points are wrongly assigned. One reason for this misassignment is the noise between $Z$ and $\tilde Z$. Therefore, to estimate the expert based only on correctly assigned labeled data points, we employ a least trimmed squares estimation-based algorithm \citep{rousseeuw2000algorithm,rousseeuw2006computing} that selectively trims potentially misassigned data points. Finally, given the parameters estimated above, we estimate the transition probability by maximizing the likelihood function based on labeled data, which is a constrained convex optimization problem. Overview of our estimation steps is summarized in Algorithm \ref{alg:overview}. 

For our theoretical analysis, we characterize the conditions under which the proposed approach performs well. It is important to note that unlabeled data does not always improve parameter estimation and may, in some cases, degrade performance, so identifying when it is beneficial is crucial. We consider the asymptotic regime where both the unlabeled data size ($N$) and the labeled data size ($n$) diverge, with the unlabeled data size diverging much faster than the other, i.e., $N,n\rightarrow\infty$ and $n/N\rightarrow 0$. We establish that, under the condition where the ``information transferable rate" from unlabeled to labeled data exceeds a specific threshold (see Assumption \ref{ass:3}), the estimators of our proposed approach converge to their true values in probability at the rate of $(n/\log n)^{-1/2}$, the parametric rate up to poly-log terms. Consequently, the predictions are asymptotically unbiased at the same rate. In Lemma \ref{lemma.1}, we further discuss an interpretable sufficient condition for this result to hold. Additionally, we show that this convergence rate is minimax optimal, and in the absence of the condition, estimation can deteriorate (see Section~\ref{subsec:thm_further}). 

The rest of this paper is organized as follows. In Section \ref{sec:model}, we introduce the noisy MoE. In Section \ref{sec:alg}, we present the semi-supervised approach for the noisy MoE. In Section \ref{sec:thm}, we establish the parametric rate of convergence previewed above and provide an outline of the proof. Section \ref{sec:review} reviews related methods, including semi-supervised and supervised approaches for MoE, and discusses them in the context of the noisy MoE. We explore the performance of our proposed approach on simulated data in Section \ref{sec:sim}. In Section \ref{sec:real}, we apply our approach to two real datasets. In Section \ref{sec:dis}, we provide a discussion.

\section{Model}\label{sec:model}

We now describe in full the {\em noisy MoE} model previewed in the introduction.  We observe two sets of random samples, $\{(x_i, y_i)\}_{i=1}^n$ and $\{ x_i \}_{i=n+1}^N$, from the following joint mixture model,
    \begin{align}
        &{\color{black}Y = \beta_{k0} + \beta_k^T X + \epsilon_k, \text{~for~} Z=k,} \label{eq:model.1} \\
        \text{and~}&X|(\tilde Z=\k) \sim N(\mu_{\k}, \Sigma_{\k}), \text{ for } k,\k = 1,\ldots, K, \label{eq:model.2}
    \end{align}
where $(X,Y) \in\mathbb R^p \times \mathbb R$, $Z$ and $\tilde Z$ are latent variables, and $n \ll N$. {\color{black}Additionally, $\epsilon_k \sim \mathrm P_{\epsilon_k}$, where $\mathrm P_{\epsilon_k}$ is a symmetric probability distribution on $\mathbb R$ such that $\mathrm E(\epsilon_k)=0$. We assume that the family of error distribution is known but its parameters are unknown.} The first set of random samples, $\{(x_i, y_i)\}_{i=1}^n$, is referred to as the labeled data and the second set, $\{ x_i \}_{i=n+1}^N$, is referred to as the unlabeled data. 

\begin{figure}[t]
\fbox{ 
  \begin{minipage}{\linewidth}
\begin{enumerate}[label={}]
\spacingset{1}
\item \underline{Variables' relation:}
    \begin{center}
        \begin{tikzpicture}[node distance=2cm]
        \node[circle,dashed,draw] at (0,0) (A) {$\tilde Z$};
        \node[circle,dashed,draw,right of=A,yshift=1cm] (B) {$Z$};
        \node[circle,draw,right of=A,yshift=-1cm] (C) {$X$};
        \node[circle,draw,right of=B,yshift=-1cm] (D) {$Y$};
        \draw[->] (A) -- (B);
        \draw[->] (A) -- (C);
        \draw[->] (B) -- (D);
        \draw[->] (C) -- (D);
        \end{tikzpicture}
    \end{center}
    \item \underline{Model:}
        \begin{equation*}
        \begin{aligned}
        &{\color{black}Y = \beta_{k0} + \beta_k^T X + \epsilon_k, \text{~for~} Z=k,}  \\
        &X|(\tilde Z=\k) \sim N(\mu_{\k}, \Sigma_{\k}), \text{ for } k,\k = 1,\ldots, K, \\
        \text{and~}&\mathrm P(Z=k|\tilde Z=\k)=\pi_{k|\tilde k}, \text{~where~} \k=\arg\max_{j}\pi_{j|\k}, \text{~for each~} \k.  
        \end{aligned}
        \end{equation*}
    \item \underline{Data:}
            \begin{equation*}
            \begin{aligned}
            &\text{Labeled data:}  &&\{(x_i, y_i)\}_{i=1}^n \\ 
            &\text{Unlabeled data:} &&\{ x_i \}_{i=n+1}^N
        \end{aligned}
        \end{equation*}
\end{enumerate}
\end{minipage}
}
\caption{A schematic of the noisy MoE model. Dashed circles represent latent variables and solid circles represent observable variables.}\label{fig:relation}
\end{figure}
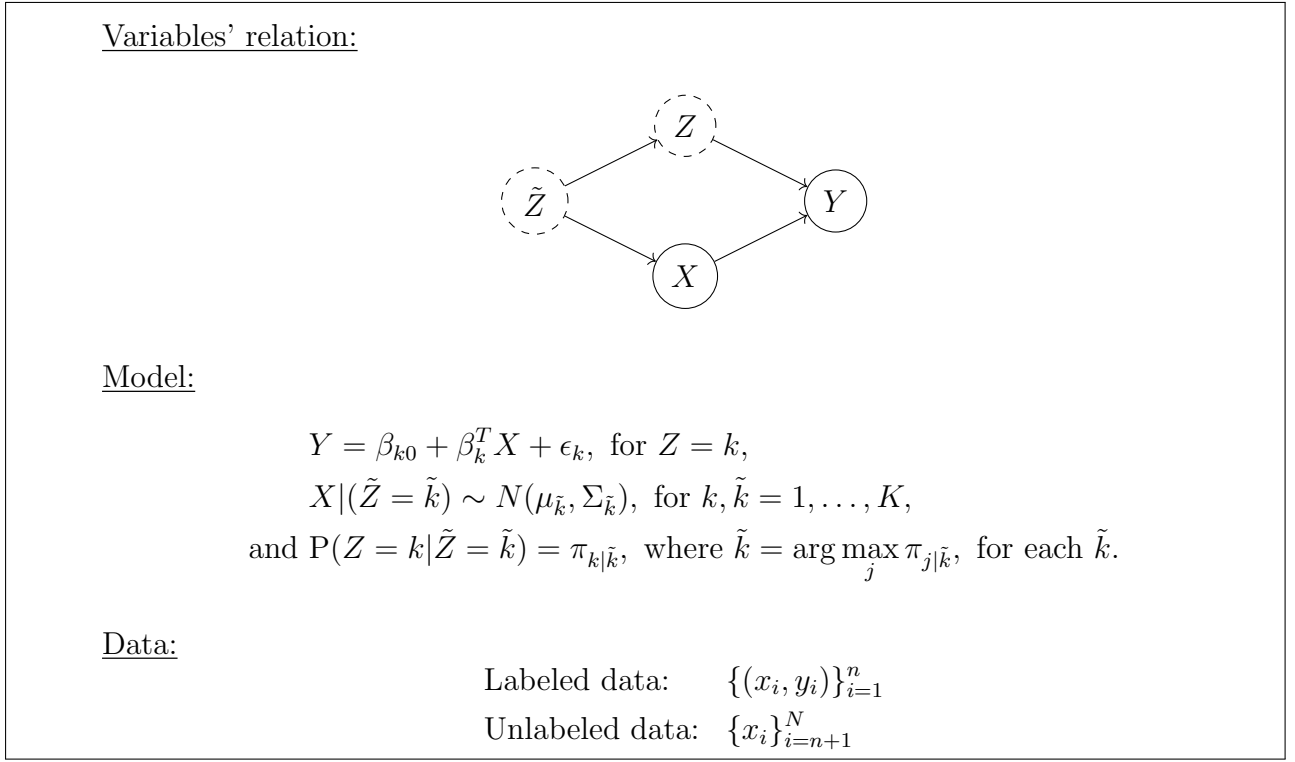

To fully specify the mixture model, we need to describe the distribution of the latent variables $(Z,\tilde Z)$ and how this relates to $X$.  We assume that $X \ind Z | \tilde Z$. The diagram illustrating the relationship between the variables is presented in Figure \ref{fig:relation}. This assumption means that the transition from the clusters defined by $\tilde Z$ to clusters defined by $Z$ shows no pattern that depends on $X$. This assumption can be reasonable in situations where the transition process is influenced by factors independent of $X$. From a model perspective, this assumption simplifies the network structure between variables, reducing the number of parameters to estimate. We discuss this assumption further in Remark \ref{rmk:assump} in the context of a data example. 
With this assumption, the joint probability of $Z$ and $\tilde Z$ given $X$ becomes  
\begin{equation}
    \mathrm P(Z=k,\tilde Z = \tilde k | X) = \mathrm P(\tilde Z = \tilde k | X) \mathrm P(Z=k | \tilde Z= \tilde k). \label{eq:latent-joint-cond-X}
\end{equation}

The conditional distribution and the conditional mean of $Y|X$ are
\begin{equation}\label{eq:dist}
    \begin{aligned}
        &\mathrm P(Y|X)  = \textstyle\sum_{k=1}^K \mathrm P(Z = k|X) \mathrm P(Y|X, Z=k) \\  
        \text{and}~ &\mathrm E(Y|X) =\textstyle\sum_{k=1}^K \mathrm P(Z = k|X) ( \beta_{k0} + \beta_k^T X ),
    \end{aligned}
\end{equation}
respectively. The model $\mathrm P(Y|X, Z=k)$ is called an expert, and $\mathrm P(Z = k|X)$ is referred to as the gate, which determines how much weight should be allocated to each expert in a given context $X$. Using \eqref{eq:latent-joint-cond-X}, the gate can be expressed as 
        \begin{equation} \label{eq:gate}
        \begin{aligned}
    \mathrm P(Z=k|X) 
    = \textstyle\sum_{\tilde k= 1 }^{K}\mathrm P(\tilde Z = \tilde k| X)\pi_{k|\tilde k } ,
        \end{aligned}
        \end{equation}
where $\pi_{k|\tilde k }:=\mathrm P(Z=k|\tilde Z=\tilde k)$ denotes the $(k,\tilde k)$-th element of the transition matrix $\Pi_{Z|\tilde Z} \in \mathbb R^{K\times K}$.
We think of the noise process of going from $\tilde Z$ to $Z$ as largely preserving the latent structure, which we express through the assumptions that $\arg\max_j\pi_{j|\k}$ is unique for each $\k$ and furthermore that $\tilde k = \arg\max_j \pi_{j|\tilde k}$.  That is, the most likely outcome when $\tilde Z=\k$ is that $Z=\k$.  This is a relaxation of the semi-supervised MoE model \eqref{eq:ssmoe} in which $\Pi_{Z|\tilde Z}=\mathbf I_K$.
We further assume that $\max_j \pi_{j|\tilde k} = \pi_{\tilde k|\tilde k}\geq 1-\delta > 0.5$ for all $\tilde k=1,\ldots, K$. This assumption suggests that the information on $\tilde{Z}$ could be helpful or transferable in identifying $Z$. 
If $\delta$ satisfies the specific condition $1-\delta > 0.5 / (1-\epsilon)$ (see Lemma \ref{lemma.1} and Assumption \ref{ass:3}), where $\epsilon$ is defined in Assumption \ref{ass:3a}, then the assumption is sufficient (but not necessary) for our proposed algorithm, introduced in the next section, to work well. For details on the necessary condition, please refer to Assumption \ref{ass:3} in Section \ref{sec:thm}.

When the parameter estimates introduced in the next section are available, the prediction is given by
\begin{equation}\label{eq:predy}
    \hat y (X) =  \textstyle\sum_{k= 1 }^{K} \textstyle\sum_{\tilde k= 1 }^{K} \hat{\mathrm P}(\tilde Z = \tilde k| X)\hat\pi_{k|\tilde k } ( \hat\beta_{k0} + \hat\beta_k^T X ),
\end{equation}
which follows from combining \eqref{eq:dist} and \eqref{eq:gate}.

\section{Algorithm}\label{sec:alg}

In this section, we present a semi-supervised estimation strategy for the parameters of the model described in the previous section. Algorithm \ref{alg:overview} provides an overview of the estimation procedure. Section~\ref{subsec:expert} details Steps 1--3, focusing on estimating experts. Section~\ref{subsec:gate} details Steps 4--5, focusing on estimating gates.  

\begin{algorithm}[t]
\caption{Algorithm overview}\label{alg:overview}
\spacingset{1}
\begin{algorithmic}
    \State Step 1: Learn the joint distribution of $(X,\tilde Z)$ from unlabeled data by fitting a Gaussian mixture model.
    \State Step 2: Assign each labeled data point to the $K$ clusters defined by $\tilde Z$, using the estimated distribution of $\tilde Z|X$. 
    \State Step 3: Estimate the experts' parameters $(\hat\beta_{0,k},\hat\beta_k)$ and {\color{black}the parameters in $\mathrm P_{\epsilon_k}$ based on} the least trimmed squares estimation method. \Comment{See Equations \ref{eq:est_trim} and \ref{eq:est_sig}.}
    \State Step 4: Estimate the $Z|\tilde Z$ transition matrix, $\Pi_{Z|\tilde Z}$.  \Comment{See Algorithm \ref{alg:gate}.}
    \State Step 5: Estimate the gate function, $\mathrm P(Z|X)$, using estimated distributions of $\tilde Z|X$ and $Z|\tilde Z$. 
  \end{algorithmic}
\end{algorithm} 

\subsection{Estimation of Experts}\label{subsec:expert}

    The assumption that $\max_j\pi_{j|k} = \pi_{k|k}\geq 1-\delta>0.5$ for all $k$ 
    suggests that there is one dominant group of $Z$ given $\tilde Z$. This means that if we condition on $\tilde Z$, at least 50\% of the samples are generated by the same expert on average. 
    However, $\tilde Z$ is the latent variable which is not observable.

    Our approach to estimating the experts is as follows: First, we identify or impute the value of the latent variable $\tilde Z$ based on $X$, denoted as $\hat s(X)$. Let us assume that $\hat s(X)$ closely approximates the true value of $\tilde Z$, so that $\max_j \mathrm P(Z = j | \hat s(X) = k) \geq 0.5$  holds for all $k$. (For this to hold asymptotically, the rigorous assumption required is presented in Assumption \ref{ass:3} in Section \ref{sec:thm}.) 
    Next, by conditioning on $\hat s(X)$, we find that over half of the samples share the same linear trend and so we can estimate this line as the mean of the experts. Below are the details of our expert estimation step. 
    
    We first estimate the distribution of $X$, which can be used to recover the unseen variable $\tilde Z$. 
    As $X$ follows the mixture of Gaussian distributions, 
    $$
        X\sim \textstyle\sum_{\tilde k=1}^{K} \mathrm P(\tilde Z= \tilde k) N(\mu_{\tilde k}, \Sigma_{\tilde k}),
    $$
    we fit the mixtures of Gaussian distributions to $\{x_i\}_{i=1}^N$ and denote the estimated parameters by adding a hat symbol ($~ \hat{}~ $) to the corresponding parameters. When $N\rightarrow \infty$, the estimated quantities approximate the true parameter values well. 

    For each $i=1,\ldots,n$, we record the cluster index $k$ to which $x_i$ is most likely to belong, denoted as $\hat s(x_i)$: 
    $$ 
        \hat s(x_i) = \arg \max_{\tilde k} \hat {\mathrm  P} (\tilde Z = \tilde k | X = x_i),
    $$
    where 
    \begin{equation}\label{eq:gate_side}
        \hat {\mathrm  P} (\tilde Z= \tilde k | X) = \frac{\hat{\mathrm P}(\tilde Z= \tilde k)N(X;\hat\mu_{\tilde k}, \hat\Sigma_{\tilde k})}{\textstyle\sum_{\tilde k=1}^{K} \hat{\mathrm P}(\tilde Z= \tilde k)N(X;\hat\mu_{\tilde k}, \hat\Sigma_{\tilde k})}.
    \end{equation}
    Define $\mathcal I_{k} = \{i : \hat s(x_i) = k \} $ to be the set of all points estimated by $\hat s(\cdot)$ to be in the $\tilde Z = k$ cluster. 
    
    From our assumption that $\mathrm P(Z = k | \hat s = k) \geq 0.5$, the majority of $y_i$ for $i\in \mathcal I_k$ are generated from the same distribution, namely, {\color{black}the distribution of $\beta_{k0} + \beta_k^T X_i + \epsilon_k$.} 
    Given that under half of the $y_i$ do not come from this underlying model, we employ a robust estimation procedure.
    Letting $\mathbf Y_{\mathcal I_k}\in\mathbb R^{|\mathcal I_k|}$ and $\mathbf X_{\mathcal I_k}\in\mathbb R^{|\mathcal I_k|\times p}$ be the labeled data response vector and predictor matrix restricted to the observations in $\mathcal I_k$,
    we estimate $\beta_{0k}$ and $\beta_{k}$ using least trimmed squares (LTS; \citealt{rousseeuw2000algorithm,rousseeuw2006computing}):
        \begin{equation}\label{eq:est_trim}
        \begin{aligned}
            & (\hat \beta_{0k}, \hat \beta_k, \hat w_k) = \arg\min_{\beta_{0k},\beta_k,w_k} \frac{ (\mathbf Y_{\mathcal I_k} - \beta_{k0} \mathbb 1  - \mathbf X_{\mathcal I_k}  \beta_k)^T \text{Diag}(w_k) (\mathbf Y_{\mathcal I_k} - \beta_{k0}\mathbb 1  - \mathbf X_{\mathcal I_k}  \beta_k) } { \|w_k\|_0 -p-1 }, \\
            & \text{subject to } w_k \in \{ 0, 1\}^{|\mathcal I_k|} ~ \text{and} ~ \|w_k\|_0 = \floor*{ \alpha \cdot ( |\mathcal I_k| + p + 1) }. \\
        \end{aligned}
        \end{equation}
    Here, $\| \cdot \|_0$ represents the number of non-zero elements in a vector, $\mathbb 1$ is a $|\mathcal I_k|$-dimensional vector of ones, and $\text{Diag}(w_k)$ is a diagonal matrix with the diagonal elements being $w_k$. 

    Non-zero elements of $w_k$ correspond to observations that were retained by LTS in estimating the parameters, so we call $\alpha$ the {\em retaining hyper-parameter}, representing the proportion of data points used in the estimation after trimming the relatively extreme residuals. To include as much data as possible, we would ideally want to take $\alpha$ for the $k$-th group to be $\mathrm P(Z = k | \hat s(X) = k)$; however, this is unknown so we take the conservative choice of $\alpha = 0.5$ in practice. 

    {\color{black}We then estimate the parameters of $\mathrm P_{\epsilon_k}$, denoted by $\theta_k\in\mathbb R^r$, where $r$ is the number of parameters in $\mathrm P_{\epsilon_k}$. The maximum likelihood estimator is applied to the retained samples after substituting the estimated values of $\beta_{0k}$ and $\beta_k$, as follows:
    \begin{equation*}\label{eq:est_sig}
    \begin{aligned}
        & \hat \theta_k = \arg\min_{\theta_k} ~ \hat w_k^T \mathbf V ( \mathbf r_{\mathcal I_k}  ;\theta_k) ,
    \end{aligned}
    \end{equation*}
    where $\mathbf r_{\mathcal I_k} = \mathbf Y_{\mathcal I_k} - \hat \beta_{k0}\mathbb 1  - \mathbf X_{\mathcal I_k}  \hat\beta_k$ and $\mathbf V ( \mathbf r_{\mathcal I_k} ;\theta_k) \in \mathbb R^{|\mathcal I_k|}$ has $i$-th component $\log \mathrm P_{\epsilon_k} ( [\mathbf r_{\mathcal I_k}]_i ;\theta_k)$. 
    For example, if $\mathrm P_{\epsilon_k}$ has the normal distribution, $N(0,\sigma_k^2)$, then 
    \begin{equation*}
    \begin{aligned}
        & \hat \sigma_{k}^2 = \frac{ (\mathbf Y_{\mathcal I_k} - \hat \beta_{k0} \mathbb 1  - \mathbf X_{\mathcal I_k}  \hat \beta_k)^T \text{Diag}(\hat w_k) (\mathbf Y_{\mathcal I_k} - \hat \beta_{k0}\mathbb 1  - \mathbf X_{\mathcal I_k}  \hat \beta_k) } { \|\hat w_k\|_0  }.
    \end{aligned}
    \end{equation*}
    }

\subsection{Estimation of Gates}\label{subsec:gate}

Recall that the gate function is
$$\mathrm P(Z = k|X) = \textstyle\sum_{\tilde k=1}^{K} \mathrm P(\tilde Z = \tilde k | X)\pi_{k|\tilde k} .$$
The first term on the right-hand side, $\mathrm P(\tilde Z = \tilde k | X)$, is estimated using \eqref{eq:gate_side}. It remains to estimate the second term $\pi_{k|\tilde k}$ for all $k,\tilde k $. 
In this section, we focus on estimating $\Pi_{Z|\tilde Z}$, the transition matrix. 

From \eqref{eq:dist}, $Y|X$ follows the mixtures distribution,
        \begin{equation*}
        \begin{aligned}
            \mathrm P(Y|X) & = \textstyle\sum_{k=1}^K \mathrm P(Z = k|X) {\color{black}\mathrm P_{\epsilon_k}(Y - \beta_{k0} - \beta_k^T X;\theta_k)},
        \end{aligned}
        \end{equation*}
and so the log-likelihood function is 
        \begin{equation*}
        \begin{aligned}
             \textstyle\sum_{i=1}^n \log \left( \textstyle\sum_{k,\tilde k} \pi_{k|\tilde k} \mathrm P(\tilde Z_i = \tilde k| X_i) {\color{black}\mathrm P_{\epsilon_k}(Y_i - \beta_{k0} - \beta_k^T X_i;\theta_k)} \right).
        \end{aligned}
        \end{equation*}
 
As the distribution of $\tilde Z|X$ and the experts' parameters are estimated in the previous section, we only need to estimate $\Pi_{Z|\tilde Z}$. We maximize the log-likelihood function while fixing the other parameters. That is, we solve the optimization problem,
        \begin{equation}\label{eq:obj_gate}
        \begin{aligned}
            \min_{\Pi_{Z|\tilde Z}} ~& \mathcal I(\Pi_{Z|\tilde Z}) =  -\textstyle\sum_{i=1}^n \log \left( \textstyle\sum_{k,\tilde k} \pi_{k|\tilde k} \hat {\mathrm P}(\tilde Z_i = \tilde k| X_i) {\color{black}\mathrm P_{\epsilon_k}(Y_i - \hat\beta_{k0} - \hat\beta_k^T X_i;\hat\theta)} \right), \\ 
            & \text{subject to~} \pi_{k|\tilde k} \geq 0 \text{~and~} \textstyle \sum_{k=1}^K \pi_{k|\tilde k} = 1.
        \end{aligned}
        \end{equation}
This is a constrained convex optimization problem, which we solve using the exponentiated gradient algorithm \citep{kivinen1997exponentiated} summarized in Algorithm \ref{alg:gate}. A detailed derivation is provided in the appendix. 
\begin{algorithm}[t]
\caption{Exponentiated gradient algorithm for solving \eqref{eq:obj_gate}}\label{alg:gate}
\spacingset{1} \begin{algorithmic}
\Require Choose a step size $s$. 
\For{$t=1,2,\ldots$}
\State (1) For all $k,\tilde k$, compute $$G_{k,\tilde k} \leftarrow \pi_{k|\tilde k}^{t} \exp \{  - s [ \triangledown \mathcal I({\hat \Pi}_{Z|\tilde Z}^t) ]_{k,\tilde k} - 1 \},$$
\State $~~~$ where $[ \triangledown \mathcal I(\hat \Pi_{Z|\tilde Z}^t) ]_{k,\tilde k}$ is defined as
        $$
            [ \triangledown \mathcal I(\Pi_{Z|\tilde Z}) ]_{k,\tilde k} = - \sum_{i=1}^n\frac{\hat {\mathrm P}(\tilde Z_i = \tilde k| X_i) {\color{black}\mathrm P_{\epsilon_k}(Y_i - \hat\beta_{k0} - \hat\beta_k^T X_i;\hat\theta)} }{\textstyle\sum_{k,\tilde k} \pi_{k|\tilde k} \hat {\mathrm P}(\tilde Z_i = \tilde k| X_i) {\color{black}\mathrm P_{\epsilon_k}(Y_i - \hat\beta_{k0} - \hat\beta_k^T X_i;\hat\theta)} }.
        $$
\State (2) Update $\hat \Pi_{Z|\tilde Z}^{t+1} \leftarrow G_{k,\tilde k} \cdot \textstyle \sum_{k=1}^K G_{k,\tilde k}$ \Comment{Rescale column to sum 1.}
\EndFor
\end{algorithmic}
\end{algorithm}

\section{Theory}\label{sec:thm}

In this section, we provide theoretical results on the operating characteristics of our proposed method. Denote the joint probability of the latent variables as $\pi_{k,\k}=\mathrm P(Z=k, \tilde Z = \k)$, and the marginals as $\pi_{k,\cdot}=\mathrm P(Z=k)$ and $\pi_{\cdot,\k}=\mathrm P(\tilde Z=\k)$, for $k,\k =1,\ldots,K$. 
As discussed in Section \ref{sec:model}, we assume that $\tilde k = \arg\max_j \pi_{j|\k}$. 
Therefore, the joint probability $\{\pi_{k,\k}: 1 \leq k,\k \leq K \}$ is a square matrix with the each row's maximum attained on the diagonal; it resembles the transition matrix in corruption models (where $Z=\tilde Z$ unless there is corruption), which have been well-studied in information theory and robust statistics \citep{van2018theory}. 

We consider an asymptotic regime where both $n$ and $N \to \infty$, but $n/N \to 0$. 
To facilitate a rigorous proof of our results, we impose the following assumptions.
    
\begin{assumption}\label{ass:1}
    Assume that the parameters in \eqref{eq:model.1} 
    are such that $\max_k |\beta_{0k}|$ and $\max_k ||\beta_k||_1$ 
    are bounded. Also, assume that the parameters in \eqref{eq:model.2} satisfy $\max_{\k}||\mu_{\k}||_1$ and $\max_{\k}||\Sigma_{\k}||_F$ are bounded, and that the marginal distribution of $X$ has $K$ distinct modes. 
\end{assumption}
\begin{assumption}\label{ass:2}
    We assume that there is no strong imbalance among the clusters in the true population distribution of $(Y,X)$ in \eqref{eq:model.1}, i.e., there exists a constant $C_0$ independent of $n$ and $N$ such that  
    \begin{align}\label{eq:A2}
    \Big(\max_{1\leq k \leq K} \pi_{k,\cdot}\Big) \Big/ \Big( \min_{1\leq k \leq K} \pi_{k,\cdot}\Big) \leq C_0~.
    \end{align}
\end{assumption}

Assumption \ref{ass:1} imposes a standard, benign condition on the model parameters and prevents trivial degeneracies in our proofs. Assumption \ref{ass:2} also poses no significant restrictive conditions and is common for theoretical study of our problem; it is essential as statistical learning in class-imbalanced data \citep{haixiang2017learning,he2009learning,lin2017clustering} requires additional methodological extensions that are not considered here.  

Before introducing other assumptions, we define the notation $s(X)$ to represent the Bayes' rule for classifying the unlabeled data $X$, i.e., $s(X) = \argmax_{\tilde k} \mathrm P(\tilde Z=\tilde k|X)$.  This is an idealized version of $\hat s(X)$ introduced in Section~\ref{sec:alg}.  Since $\tilde Z$ is an unobservable variable, the Bayes rule $s(X)$ provides the best possible approximation based on $X$. Our next assumption is directly related to the operating characteristics of our proposed method. 

\begin{assumption}\label{ass:3}
Assume that
\begin{align}\label{eq:A1}
\gamma_{0} := \min_{1\leq\k \leq K}\mathrm P \big(Z = \k \big| s(X)=\k \big) > 0.5.
\end{align}
\end{assumption}

To understand the impact of Assumption \ref{ass:3}, first note that $\gamma_0$ can be interpreted as the minimum probability of association with the dominant expert, given a cluster defined by the Bayes rule.
Therefore, $\gamma_0$ is an important characteristic of the model, reflecting the degree of usefulness of the information presented in the unlabeled data for estimating the experts' parameters in the labeled data. 
Intuitively, as $\gamma_0$ increases, estimating the parameters in model \eqref{eq:model.1} becomes easier as the information in the unlabeled data can be well-leveraged to improve estimation accuracy of the experts' parameters in the labeled data. Since we run an LTS-based algorithm with a retaining parameter being at least $0.5$, we require $\gamma_0 > 0.5$ in \eqref{eq:A1} for our proposed algorithm not to break down. We later explain the role of $\gamma_0$ through more interpretable conditions.

We now present our main result. To facilitate an easier proof of our results, we consider working with bounded covariates for the labeled data. For this purpose, we screen out labeled data points with  $\|x\|_2$ large. As $\{x_i: 1\leq i \leq n\}$ are from \eqref{eq:model.2}, for any $q>0$, the $(1-q)$th quantile of $\{\|x_i\|_2: 1 \leq i \leq n\}$ is bounded in probability as $n \to \infty$. Thus, the screening process would not decrease the effective sample size of labeled data by much.   
We first show that Assumption \ref{ass:3} can be extended for bounded covariates without loss of generality. Next, we show that if the noise distribution in \eqref{eq:A1.3} has bounded support then the proposed algorithm, when run on the aforementioned screened version of the labeled data, efficiently estimates the expert means.

{\color{black}
\begin{theorem}\label{thm.1} In the noisy MoE model of \eqref{eq:model.1}--\eqref{eq:model.2}, under Assumption \ref{ass:3}, there exists $M^\ast$ such that for all $M \geq M^\ast$,
\begin{align}\label{eq:ass:3_variant}
    \gamma^\ast_{0} = \min_{1\leq\k \leq K} \mathrm P \big(Z = \k \big|  s(X)=\k, \| X \|\leq M \big) > 0.5.
\end{align} 
Under Assumptions \ref{ass:1}--\ref{ass:3}, if the noise distribution in \eqref{eq:A1.3} has bounded support, and $\min_{k\neq l}| \beta_{0k} - \beta_{0l}|$ is large, then using our proposed method with retaining hyper-parameter $\alpha \in [0.5,\gamma^\ast_{0})$ on screened labeled data, the estimation errors of the experts' means can be controlled:  
\begin{align*}
     \max_{1\leq k \leq K} \;  \big\Vert (\hat{\beta}_{0k},\hat{\beta}_{k})-(\beta_{0k},\beta_{k}) \big\Vert_1 = o_p(c_n) \text{ as } n \to \infty \text{, where, } c_n=(n/ \log n)^{-1/2}. 
\end{align*}
\end{theorem}
}

Theorem~\ref{thm.1} shows that our proposed method given in \eqref{eq:est_trim} for estimating the expert means  can obtain the parametric rate up to poly-log terms. 

{\color{black}
\begin{remark}
For the theoretical results, for simplicity, we have assumed that $\theta_k$ in \eqref{eq:model.1} are known. In LTS, unbiased estimation beyond the means can be challenging. However, since we know the error distribution, we can efficiently estimate $\theta_k$ by examining the different quantiles of the residuals from the retained samples in LTS and using plug-in estimates for the expert means. 
\end{remark}
}

As described in our methodology section, once we get estimates of the coefficients via LTS, we use them to estimate the unknown $\Pi_{Z|\tilde Z}$ matrix. Using those estimates, we predict the responses based on the $X$ values of future observations.  As we have $\sqrt{n}$-consistent estimates of the expert means, the $\Pi_{Z|\tilde Z}$ matrix can be well estimated. The following result shows that the error of our predictor $\hat{y}(X)$ is well controlled in the long run. As such it is asymptotically unbiased and has bounded $L_1$ risk.

\begin{corollary}\label{cor.1}
    Under the conditions of Theorem~\ref{thm.1}, the error associated with our proposed predictor $\hat{y}(X)$ \eqref{eq:predy} of the future response $Y$ based on its covariate value $X$ is 
    $$ \mathrm E[Y-\hat{y}(X)\big|X]=o_p(c_n) \text{ and } \mathrm E\left[|Y-\hat{y}(X)|\big|X\right]=O_p(1) \text{ as } n \to \infty.$$ 
\end{corollary}

\subsection{Overview of the Proof of Theorem~\ref{thm.1}} 
A detailed proof of Theorem~\ref{thm.1} and Corollary~\ref{cor.1} is presented in the appendix. Here, we briefly present the heuristic idea behind the proof. The main idea in the proof is that for estimating the mixture regression coefficients in \eqref{eq:model.1} we can essentially restrict ourselves to the set of observations in $\{1,\ldots,n\}$ with $\| x_i\|_2\leq M$ for which our clustering algorithm $\hat s$ classifies the corresponding $X$ observation to cluster $\l$,
$$\hat{\mathcal{S}}(\l)=\{1\leq i \leq n: \hat s(x_i)=\l \text{~and~} \| x_i \| \leq M \},$$ 
and then conduct least trimmed square (LTS) regressions separately for each $\SS(\l)$ as $\l$ varies over 
$\{1,\ldots,K\}$.

In the set $\hat{\mathcal{S}}(\l)$ with high probability we show that more than a fraction $\alpha$ of the points are associated with only one group in the labeled model of \eqref{eq:model.1}, which is $\l$.  

For each $\l \in \{1,\ldots,K\}$ consider the following two subsets on $\{1,\ldots,n\}$: 
$$\tilde{\mathcal{Z}}(\l)=\big\{1\leq i \leq n: \tilde Z_i=\l  \text{~and~} \| x_i \| \leq M  \big\} \text{ and } \mathcal{Z}(\l)=\big\{1\leq i \leq n: Z_i=\l  \text{~and~} \| x_i \| \leq M  \big\}.$$
Note that, unlike $\hat{\mathcal{S}}(\l)$ we do not observe $\tilde{\mathcal{Z}}(\l)$ or ${\mathcal{Z}}(\l)$ because they depend on latent variables.
 
We next decompose $\SS(\l)$ into two non-overlapping sets. \begin{itemize}
\item [(I)] $\mathcal{A}_1(\l)=[\SS  \cap  \ZZ](\l)$: all observations in $\SS(\l)$, for which the $Y$ values conditioned on the $X$ values are from the expert indexed by $\l$ in \eqref{eq:model.1}. 
\item [(II)] $\mathcal{A}_2(\l)=[\SS  \cap  \ZZ^C](\l)$: all observations in $\SS(\l)$, for which the $Y$ values conditioned on the $X$ values are \textit{not} from the expert indexed by $\l$ in \eqref{eq:model.1}. 
\end{itemize}

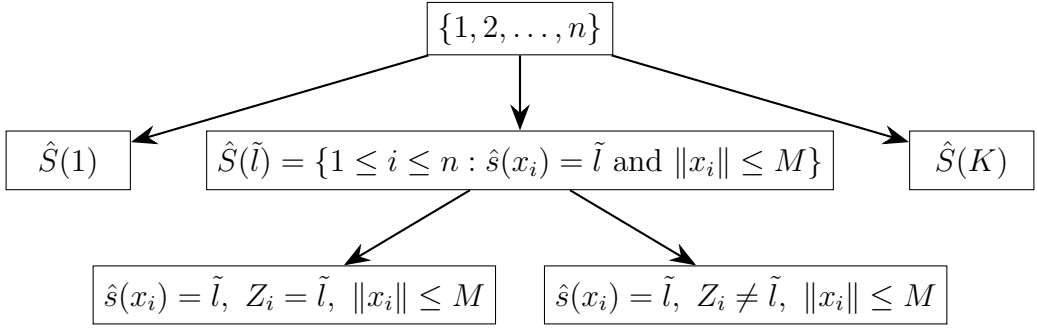
\begin{figure}[!h]
\centering
\begin{tikzpicture}[
    node distance=1cm and 1cm,
    box/.style={draw, rectangle, minimum height=1.5em, minimum width=4em, text centered},
    arrow/.style={-{Stealth[scale=1.5]}, thick},
    mynode/.style={draw, rectangle, minimum height=1.5em, minimum width=6em, text centered}
]

\node[box] (Top) {$\{1,2,\ldots,n\}$};
\node[box, below=of Top] (Si) {$\hat{S}(\tilde l) = \{1 \leq i \leq n: \hat{s}(x_i) = \tilde{l}  \text{~and~} \| x_i \| \leq M  \}$}; 
\node[box, left=of Si] (S1) {$\hat{S}(1)$};
\node[box, right=of Si] (Sk) {$\hat{S}(K)$};

\node[mynode, below=of Si, xshift=-3cm] (Left) {$\hat{s}(x_i) = \tilde{l},~ Z_i = \tilde l,~\|x_i\| \leq M$};
\node[mynode, below=of Si, xshift=3cm] (Right) {$\hat{s}(x_i) = \tilde{l},~ Z_i \neq \tilde l,~\|x_i\| \leq M$};

\draw[arrow] (Top) -- (S1);
\draw[arrow] (Top) -- (Si);
\draw[arrow] (Top) -- (Sk);
\draw[arrow] (Si) -- (Left);
\draw[arrow] (Si) -- (Right);
\end{tikzpicture}
\caption{Schematic of the proof of Theorem~\ref{thm.1} 
}\label{fig.theory}
\end{figure}

In Figure~\ref{fig.theory}, we provide a schematic demonstrating this decomposition through a decision tree. It shows that we start with the partitions of the labeled data based on the classifier $\hat s$ trained on $N-n$ unlabeled observations $x_{n+1},\ldots,x_{N}$.  Each of the partitions is then sub-divided into leaves. 
We next calculate the asymptotic sizes of each of the above sets as $n \to \infty$. We show that under \eqref{eq:A1}  the asymptotic share of $\SS \cap \ZZ$ in $\SS$, i.e., $|\mathcal{A}_1(\l)|/|\SS(\l)|$ exceeds $\alpha$ for all $\l$. Thus, running LTS with the retaining hyper-parameter $\alpha$ on each $\SS(\l)$ will approximately provide us the mean of observations from $|\mathcal{A}_1(\tilde l)|$ only, which yield asymptotically efficient estimates of $\beta_{\l}$ as $n \to \infty$.

The primary new theoretical ingredients in our proof is rigorously showing that the share of $\mathcal{A}_1(\l)$ exceeds $\alpha$ in the novel settings of the noisy MoE model, \eqref{eq:model.1}-\eqref{eq:model.2}. Thereafter, to arrive at the conclusion of Theorem~\ref{thm.1} we use proof techniques from the existing literature \citep{zuo2022asymptotics,zuo2023least,vivsek2006least} that are used in studying the breakpoint analysis and asymptotic properties of LTS. We provide these proof details in the appendix. We end this section by providing a heuristic calculation on the sizes of sets in (I) and (II).    

Note that, the expected value of the sets are given by, 
$$\mathrm E|\mathcal{A}_1(\l)|=n \,\P(\hat{s}(X)=\l,Z=\l, \| X \|\leq M) \approx n \,\P(s(X)=\l,Z=\l, \| X \|\leq M) :=I_{1,n},$$
where, the approximation is based on the fact that the empirical performance of the classifier $\hat{s}(X)$ (which is trained on $N-n$ unlabeled data) converges to the accuracy of the Bayes' rule. We have provided rigorous analysis of this approximation in the appendix. By similar calculations, we show that
\begin{align*}
&I_{2,n}:=\mathrm E|\mathcal{A}_2(\l)|=\mathrm E|\SS(\l)| - \mathrm E|\mathcal{A}_1(\l)|, \\
\text{where~}& \mathrm E|\SS(\l)| = n \,\P(\hat{s}(X)=\l , \| X \|\leq M) \approx n \,\P(s(X)=\l, \| X \|\leq M ).
\end{align*}
Next, note that for all $n$, 
$$I_{1,n}/(I_{1,n}+I_{2,n}) = \P \big( Z = \l \big| s(X)=\tilde l, \| X \|\leq M \big) > \alpha \geq 0.5 $$ 
by Assumption \ref{ass:3}. 
 
In the appendix, we conduct second-order calculations and evaluate the variance of the sizes of these sets. We show that asymptotically there is concentration around the mean and $I_{1,n}/(I_{1,n}+I_{2,n})> \alpha$ ensures 
$\P(|\mathcal{A}_1(\l)|/|\SS(\l)|>\alpha)\geq 1-O_p(n^{-1})$.
On the set $\{|\mathcal{A}_1(\l)|/|\SS(\l)|>\alpha\}$, following the existing literature on LTS \citep{zuo2022asymptotics}, we show that the experts' means converges at the rate $c_n^{-1}$, completing the proof. 
 
\subsection{Explanation of the Main Result}

To gain a deeper understanding of the condition used in Theorem~\ref{thm.1}, where the retaining hyper-parameter $\alpha$ is controlled below $\gamma_0$, we present the following more interpretable assumptions on the model parameter that would ensure this condition holds. These assumptions are sufficient but not necessary for Assumption~\ref{ass:3}. That is, Assumption~\ref{ass:3} is a less restrictive requirement than the assumptions discussed below.

For $1\leq \k  \leq K$, consider the following two assessment criteria on the Bayes rule $s(\cdot)$ in \eqref{eq:model.2}.  The first is a conditional measure of its accuracy whereas the second is a relative change criterion on the marginals and hence a marginal discrepancy measure:  
$$m(\k)=\P\big( s(X)=\k \big|\tilde Z=\k \big) \text{ and } r(\k)=\{\P( s(X)=\k)-\P(\tilde Z=\k)\}/\P(\tilde Z=\k).$$ 

Assume the conditional accuracy measure of the Bayes rule is high and the marginal discrepancies are low as follows. 
\begin{assumption}\label{ass:3a}
There exists $\epsilon \in [0,1)$ for which:   
\begin{align}\label{eq:A1.2}
\inf_{\k=1,\ldots,K} \quad \frac{m(\k)}{1+r(\k)} \geq 1-\epsilon~. 
\end{align} 
\end{assumption}

 Note that $\epsilon$ is one of the constituents in our asymptotic analysis that reflects the strength of the information in the auxiliary unlabeled data. If $\epsilon$ is very small, then the population distribution of unlabeled data has well-separated sub-populations and so, we can classify them into different groups with high accuracy. However, this does not guarantee that an algorithm will be able to predict the response $Y$ well unless there is strong contiguity between the clustering structure in $X$ and the clustering structure in $Y|X$. Our next assumption is on the contiguity between these two set of groups. 

\begin{assumption}\label{ass:3b}
There exists $\delta \in [0,1)$ such that, for each $\k \in \{1,\ldots,K\}$, we have: 
\begin{align}\label{eq:A1.3}
\max_{1\leq k \leq K} \pi_{k|\k} \geq 1-\delta~.
\end{align}
\end{assumption}
In the corruption model discussed before Assumption \ref{ass:1}, \eqref{eq:A1.3} implies that the corruption proportion from any cluster in $X$ to any cluster in $Y|X$ is at most $\epsilon$.

In conjunction with Theorem~\ref{thm.1}, the following lemma shows that a sufficient condition for our method to work accurately is that the
\text{retained proportion of data} $\alpha$ in our proposed method  
be less than the product of the Bayes classification accuracy of auxiliary data in \eqref{eq:A1.2} and the \text{non-corruption probability} in \eqref{eq:A1.3}.

\begin{lemma}\label{lemma.1}
Under Assumptions \ref{ass:3a} and \ref{ass:3b},  $\gamma_0 \geq (1-\epsilon) (1-\delta)$ in the noisy MoE model, \eqref{eq:model.1}--\eqref{eq:model.2}. 
\end{lemma}

\subsection{Further Results}\label{subsec:thm_further}
While Theorem~\ref{thm.1} provides a rigorous proof on the near-parametric rate of convergence of the parameter estimates produced by our proposed method, the convergence rate of $\sqrt{n}$ does not depend on $N$, and so, does not improve as $(N-n)/n$ increases. However, this is not a limitation of the proposed method, but rather an artifact of our asymptotic setup. In fact, we can show that the convergence rate of our proposed estimator in Theorem~\ref{thm.1} is minimax optimal, up to poly-logarithmic terms in our set-up. As $n \to \infty$, consider the class $\mathcal{B}_n$  of all estimators $\hat B_n=[(\hat{\beta}_{01,n},\hat{\beta}_{1,n}), \ldots, (\hat{\beta}_{0K,n},\hat{\beta}_{K,n}) ] $ of the experts' means  $\{ (\beta_{0k},\beta_{k}) : k=1,\ldots,K\}$.  

\begin{lemma}\label{lemma.f1}
In the noisy MoE model, \eqref{eq:model.1}-\eqref{eq:model.2}, under Assumptions \ref{ass:2} and \ref{ass:3}, for any sequence $a_n$ such that $\lim_{n \to \infty} n\, a_n^{-2} =0$, the estimation errors of the experts' means $(\beta_{0k},\beta_{k})$ 
in the labeled data satisfies:   
\begin{align*}
    \liminf_{n \to \infty} \; \;  \; \min_{\hat B_n \in \mathcal{B}_n} \quad \P\bigg[ \max_{ \substack{ \beta_{01}, \ldots, \beta_{0K} \in \mathbb{R} \\
{\beta}_1, \ldots, {\beta}_K \in \mathbb{R}^p } } \;\; \bigg\{a_n \cdot \max_{1\leq k \leq K} \; \big\Vert (\hat{\beta}_{0k,n},\hat{\beta}_{k,n}) - (\beta_{0k},\beta_{k}) \big\Vert_1 \bigg\}\geq C \bigg]=1. 
\end{align*}
for any $C>0$.
\end{lemma}

In this context, we note two contributions of auxiliary information in the unlabeled data that is reflected in our asymptotic analysis: 
(a) it helps in segmenting the response prediction problem in the labeled data into sub-populations based on the covariate value. These sub-populations are not entirely homogeneous but have a significant fraction of members who have a fixed linear relationship  between the response and the covariates. The remaining fraction can be treated as corruptions and can be weeded out using LTS; 
(b) due to the presence of auxiliary data, we convert the complex mixture of experts problem on the labeled data to a regression problem that is well studied in the robust statistics literature. So, although the impact of side information in the unlabeled data is not directly reflected in the convergence rate, its presence helps to convert the problem into $K$ separate robust regression problems that are theoretically tractable and asymptotically rate-optimal.  

In contrast, methods based solely on labeled data, such as mixtures of experts or related methods that involve the {\em expectation-maximization} (EM) or {\em minorize-maximization} algorithms to train the parameters, have an extremely complicated optimization landscape \citep{Balakrishnan2017}. Existing theoretical studies on the convergence rates of such methods often involve delicate assumptions about appropriate initialization of the algorithm. \citet{Kwon2020} use tensor method to initialize EM with a point that is close to the solution and need to involve sample splitting for convergence analysis of the algorithm. Recently, \citet{Wang2024} provided finite sample bounds on mixture of regressions in regimes where predictors are random and independent of the latent variable; thus, these results do not directly extend to mixture of experts models. 

The proposed method is not, however, suitable if $\gamma_0$ is small and there is high corruption. Ideally, we would not set $\alpha$ to a low value, particularly less than $0.5$, as using a low $\alpha$ means discarding many data points, leading to higher estimation error of the mixture coefficients in finite samples. Unfortunately, this attribute is masked in our asymptotic analysis. Using the proposed algorithm with a reasonably low $\alpha$ value can lead to other issues in high-corruption settings. The following result shows that the proposed method produces inconsistent estimates if $\gamma_0$ is small. In such cases, mixtures of experts based solely on labeled data would outperform our proposed approach.

\begin{lemma}\label{lem.3}
If $\alpha > \gamma_0$, for any $C > 0$ there exists expert means $\{{\beta}_{0k},{\beta}_{k}: k=1,\ldots,K\}$ such that their estimates $\{\hat{\beta}_{0k,n},\hat{\beta}_{k,n}: k=1,\ldots,K\}$ produced by our algorithm follow
$$\liminf_{n \to \infty} \P\bigg( \max_{1 \leq k \leq K}\big \Vert  (\hat{\beta}_{0k,n},\hat{\beta}_{k,n}) -({\beta}_{0k,n},{\beta}_{k,n})\big \Vert_1 > C \bigg) \to 1 \text{ as } n \to \infty.$$
\end{lemma}

\section{Related Methods: Mixture of Experts}\label{sec:review}
In this section, we review semi-supervised and supervised approaches for the mixture of experts in the respective subsections. Additionally, we discuss these approaches in the context of the noisy MoE model. 

\subsection{Semi-supervised Mixture of Experts}\label{subsec:lit_method}

We introduce the semi-supervised approach, as described by \cite{moore2001fuzzy,jackson2017improved,hieronymi2017olci}, within a generalized framework in the context of the model \eqref{eq:ssmoe}. 
The model \eqref{eq:ssmoe} is a special case of the noisy MoE model, \eqref{eq:model.1}-\eqref{eq:model.2}, when $\mathrm P (Z=\tilde Z)=1$. Therefore, the semi-supervised MoE approach can be thought of as estimating based on the clusters defined by $\tilde Z$, while our approach is based on the clusters defined by $Z$. The semi-supervised MoE estimates $\mathrm E(Y|X)$ by
$$
    \hat y^{SS}(X) =\textstyle\sum_{k=1}^{K} \hat{\mathrm P}(\tilde Z = k|X) ( \hat{\beta}^{SS}_{k0} + X^T\hat{\beta}^{SS}_k ),
$$
where $\hat{\mathrm P}(\tilde Z = k|X)$ is estimated as in \eqref{eq:gate_side}, in the same unsupervised manner as we do in our proposed algorithm, and 
        \begin{equation*}
        \begin{aligned}
            & (\hat \beta^{SS}_{0k}, \hat \beta^{SS}_k) = \arg\min_{\beta_{0k},\beta_k} \sum_{i\in \mathcal I_k}  (Y_i - \beta_{k0} - \beta_k^T X_i)^2 . \\
        \end{aligned}
        \end{equation*}

To discuss its estimate in the context of the noisy MoE, \eqref{eq:model.1}--\eqref{eq:model.2}, we express the conditional distribution and the conditional mean of $Y|X$ from the noisy MoE, as defined in \eqref{eq:dist}, in terms of $\mathrm P(\tilde Z|X)$:
\begin{equation*}
    \begin{aligned}
        \mathrm E(Y|X) &=\textstyle\sum_{k=1}^K \mathrm P(Z = k|X) ( \beta_{k0} + \beta_k^T X ) \\
        &=\textstyle\sum_{\tilde k=1}^{K} \mathrm P(\tilde Z = \tilde k | X) ( \textstyle\sum_{k=1}^K  \pi_{k|\tilde k} \beta_{k0} + X^T \textstyle\sum_{k=1}^K\pi_{k|\tilde k}\beta_k ) , \\
        \text{and}~ \mathrm P(Y|X)  &= \textstyle\sum_{k=1}^K \mathrm P(Z = k|X) \mathrm P(Y|X, Z=k) \\
                & = \textstyle\sum_{\tilde k=1}^{K} \left\{ \mathrm P(\tilde Z = \tilde k | X)  \cdot \textstyle\sum_{k=1}^K \pi_{k|\tilde k} \mathrm P(Y|X, Z=k) \right\} . 
    \end{aligned}
\end{equation*}
As the conditional mean of the noisy MoE model, $\mathrm E(Y|X)$, can be well-expressed in terms of $\tilde Z$-clusters, the estimates $\hat \beta^{SS}_{0k}$ and $\hat \beta^{SS}_k$ can be thought of as the estimators for $ \textstyle\sum_{k=1}^K  \pi_{k|\tilde k} \beta_{k0}$ and $\textstyle\sum_{k=1}^K\pi_{k|\tilde k}\beta_k$, respectively. 

If we try to understand the noisy MoE model, $\mathrm P(Y|X)$, in terms of $\tilde Z$-clusters, the gate is $\mathrm P(\tilde Z|X)$ and each expert is the mixture of distributions, $\textstyle\sum_{k=1}^K \pi_{k|\tilde k} \mathrm P(Y|X, Z=k)$. The experts defined by $\tilde Z$-clusters generally have a larger variance than the experts defined by $Z$-clusters because the former includes additional variability introduced by the mixing. 

Therefore, if the semi-supervised MoE approach is used for the noisy MoE problem, it can be interpreted as targeting the correct mean, but the estimation of experts is based on experts with generally larger variance. This observation is noted in our simulation study in Section \ref{sec:sim}.

\subsection{(Supervised) Mixture of Experts}

In this subsection, we introduce the typical supervised approach for the mixture of experts, which utilizes only labeled data for fitting. 

The mixtures of expert model by \cite{jacobs1991adaptive} is defined as the following mixture of Gaussians distribution:
\begin{equation}\label{eq:moe}
    \textstyle\sum_{k=1}^K w_k(X;\{ \alpha_{0k},\alpha_k \}_{k=1}^K ) \mathrm P (Y|X,Z=k),
\end{equation}
where $Y|(X,Z=k) \sim N(\beta_{k0} + \beta_k^T X, \sigma_k^2)$ for $k = 1,\ldots, K$ and $w_k$ is defined by the softmax function:
$$
    w_k^{\text{line}}(X; \{ \alpha_{0k},\alpha_k \}_{k=1}^K ) = \frac{\exp(\alpha_{0k} + X^T\alpha_k)}{\textstyle\sum_{k=1}^K\exp(\alpha_{0k} +X^T\alpha_k)} .
$$
 
\cite{jordan1994hierarchical} considered the EM algorithm \citep{dempster1977maximum} for maximizing the log-likelihood function. The derivation of estimators using the EM algorithm is not detailed here, as it is well-known.

If we denote the estimated parameters by adding a hat symbol with the superscript $S$ ($~ \hat{}^{~S} ~ $), the response is predicted by
$$
\hat y^S(X) = \textstyle\sum_{k=1}^K w^{\text{line}}_k(X;\{ \hat\alpha^{S}_{0k},\hat\alpha^S_k \}_{k=1}^K ) ( \hat{\beta}^{S}_{k0} + X^T\hat{\beta}^{S}_k ).
$$

The mixture of expert model \eqref{eq:moe} and the noisy MoE \eqref{eq:model.1}--\eqref{eq:model.2} have different gate functions. If this approach is used to fit the noisy mixture of experts, it can lead to model misspecification. 

To bridge this gap, we can consider an alternate gate function proposed by \cite{xu1994alternative}, which uses the posterior probabilities of a mixture of Gaussians. We slightly generalize their gate function by considering a softmax function with a quadratic term in $X$ in place of $w_k^{\text{line}}$ for ease of computation:
$$
    w_k^{\text{quad}}(X; \{ \alpha_{0k},\alpha_k,\gamma_{\cdot,\cdot,k} \}_{k=1}^K ) = \frac{\exp(\alpha_{0k} + X^T\alpha_k + \sum_{ij}\gamma_{ijk} X_i X_j)}{\textstyle\sum_{k=1}^K\exp(\alpha_{0k} +X^T\alpha_k+ \sum_{ij}\gamma_{ijk} X_i X_j))},
$$
where $X_i$ is the $i$-th element of $X$.
Although this does not completely solve the issue but provides some relaxation since the gate function of the noisy MoE \eqref{eq:gate} is a linear combination of softmax functions with a quadratic term in $X$.

\section{Simulation}\label{sec:sim}
In this section, we compare the performance of four estimators in simulation: the semi-supervised noisy mixture of experts (NoisySS), the semi-supervised mixture of experts (MoESS), and the two versions of the supervised mixture of experts discussed in the previous section, with linear gates (MoEline) and with quadratic gates (MoEquad). We conduct these comparisons in two scenarios: one with a fixed labeled data size  and varying corruption levels, and the other with varying labeled data sizes and a fixed corruption level, where the corruption level is defined as $100\cdot (1-\textstyle\sum_{k} \mathrm P(\tilde Z = k, Z=k) )\%$. 

We consider a setting where $\{ (x_i, y_i) \}_{i=1}^n$ and $ \{ x_i \}_{i=n+1}^N$ are random samples from the model 
\begin{equation*}
    \begin{aligned}
        &Y|X,Z=k \sim N(\beta_{k0} + \beta_k^T X, \sigma_k^2) && \\ 
        &X | \tilde Z=\tilde k \sim N(\mu_{\tilde k}, \Sigma_{\tilde k}) && \text{for } k,\tilde k = 1,\ldots, K,
    \end{aligned}
\end{equation*}
where $(X,Y)\in\mathbb R^p \times \mathbb R$, $X \ind Z | \tilde Z $ and 
\begin{equation*}
\begin{aligned}
    \Pi_{Z|\tilde Z} 
    = & \begin{bmatrix}
        p_0 & \frac{1-p_0}{K-1} & \cdots & \frac{1-p_0}{K-1} \\
        \frac{1-p_0}{K-1} & p_0 & \cdots & \frac{1-p_0}{K-1} \\
        \vdots & \vdots & \ddots & \vdots \\
        \frac{1-p_0}{K-1} & \frac{1-p_0}{K-1} & \cdots & p_0
    \end{bmatrix}.
        \end{aligned}
        \end{equation*}
This transition matrix describes a simple corruption process in which the noise changes the latent cluster assignment with probability $1-p_0$, and when it does, it chooses a wrong cluster uniformly at random from the $K-1$ wrong choices.

In our simulations, we set $K = 10$ and $p = 3$. We assume the distribution of $X$ is known by setting $N=\infty$. For the parameters related to the distribution of $X$, we set $\mathrm P(\tilde Z = \tilde k) = 0.1$ for $\tilde k=1,\ldots,K$; additionally, each entry of $\mu_{\tilde k} \in \mathbb R^p$ is set to $-3 + (\tilde k-1)6/9$. We also specify $\Sigma_{\tilde k} = R_{\tilde k} D_{\tilde k} R_{\tilde k}^T$, where the entries in $R_{\tilde k} \in \mathbb R^{p\times p}$ are generated from $N(0,1)$ and then orthogonalized such that $R_{\tilde k}^T R_{\tilde k} = R_{\tilde k} R_{\tilde k}^T = \mathbf I_p$. The matrix $D_{\tilde k} \in \mathbb R^{p\times p}$ is a diagonal matrix with diagonal entries generated from $U[0.005,0.05]$. For the parameters related to the experts, we set each entry of $( \beta_{0k}, \beta_{k}^T )^T$ to be $-1 + (k-1)2/9$ and set $\sigma_k = 0.1$.

We compare different estimators using two measures. The first measure is the mean squared error (MSE) of the experts' estimators: 
$$\text{MSE}(\hat\beta,\beta) = \textstyle\sum_{k=1}^K \| ( \beta_{0k}, \hat\beta_{k}^T )^T - (\beta_{0k}, \beta_k^T)^T \|^2 / (p+1).$$
For the mixture of experts, we rearrange the indices of the $\hat\beta_k$ estimates to minimize the MSE using the Hungarian algorithm \citep{kuhn1955hungarian,R_RcppHungarian}, i.e., $\text{MSE}(\hat\beta,\beta) = \min_\tau \textstyle\sum_{k=1}^K \|\hat\beta_{\tau(k)} - \beta_k \| / (p+1)$ where $\tau(k)$ is a permutation of indices. The second measure is the relative prediction error (RPE) on the test data with a sample size of $20,000$, which is the prediction error relative to the true expected value:
$$\text{RPE}(\hat y^{\text{test}}) = \textstyle\sum_{i=1}^{20,000}  ( y_i^{\text{test}} - \hat y_i^{\text{test}} ) ^2 / \textstyle\sum_{i=1}^{20,000} (y_i^{\text{test}} - \mathrm E(Y|x_i^{\text{test}}) )^2 , $$
where $\mathrm E(Y|x_i) = \textstyle\sum_{k=1}^K \mathrm P(Z_i = k|x_i) ( \beta_{k0} + \beta_k^T x_i )$.

In the following subsections, we present the performance of the algorithms described above across a broad range of regimes. To complement the theoretical results from the previous section, these numerical experiments demonstrate the performance of our proposed LTS-based algorithm in finite samples and under unbounded noise distributions.

\subsection{Varying Corruption Levels}
We generate a labeled training set of size $n=2,000$ and a test set of size $20,000$. We try seven different $p_0$ values: $0.4$, $0.5$, $0.6$, $0.7$, $0.8$, $0.9$, and $1.0$. The corresponding corruption levels (\%) are 60, 50, 40, 30, 20, 10, and 0, respectively. We then compute each estimator's MSE and RPE. 

\begin{figure}[h!t]
    \centering
    \begin{subfigure}[b]{0.48\textwidth}
        \centering
        \includegraphics[width=\textwidth]{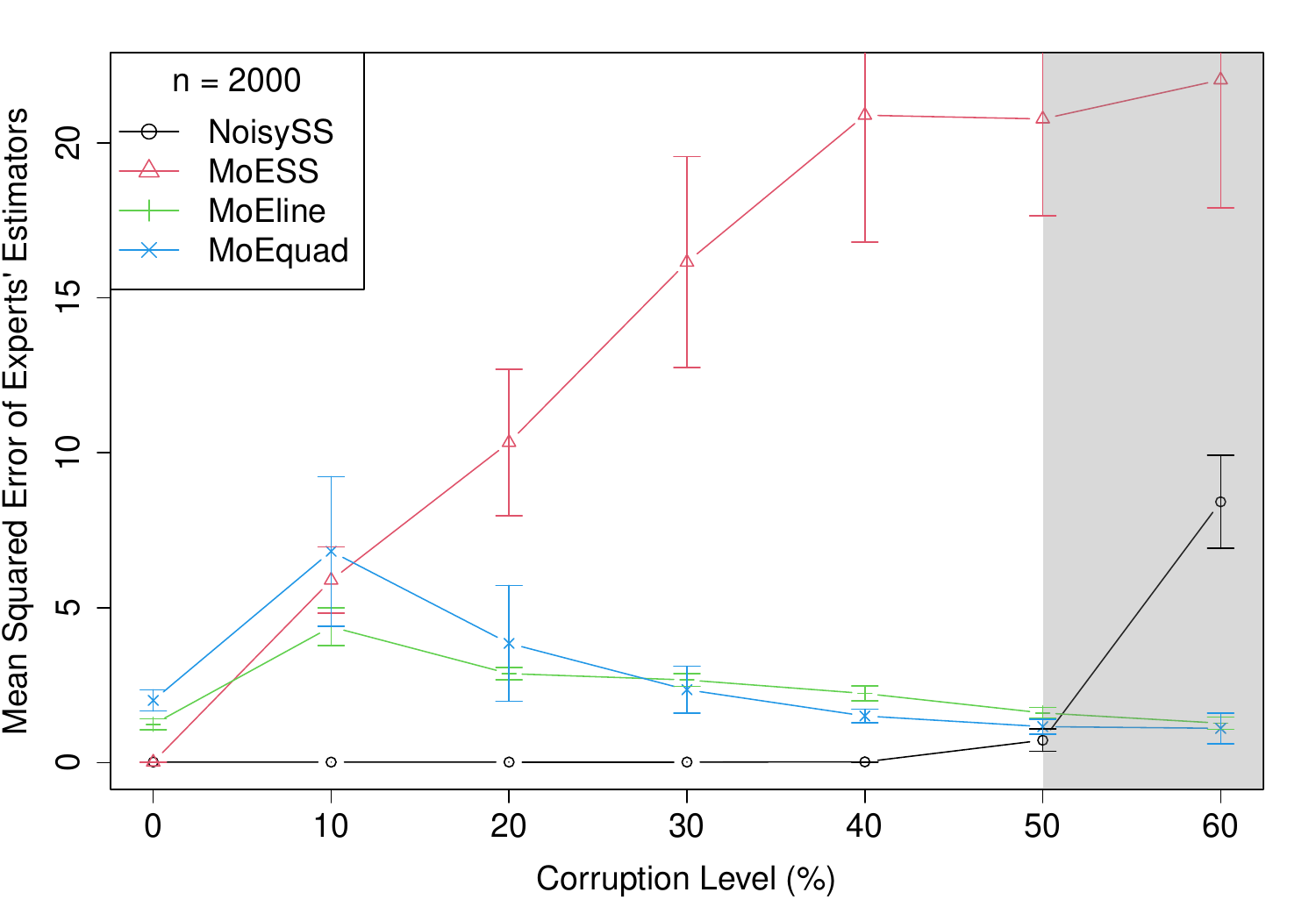}
        \caption{Average of MSE Comparison}
        \label{fig:fixn_beta}
    \end{subfigure}
    \hfill
    \begin{subfigure}[b]{0.48\textwidth}
        \centering
        \includegraphics[width=\textwidth]{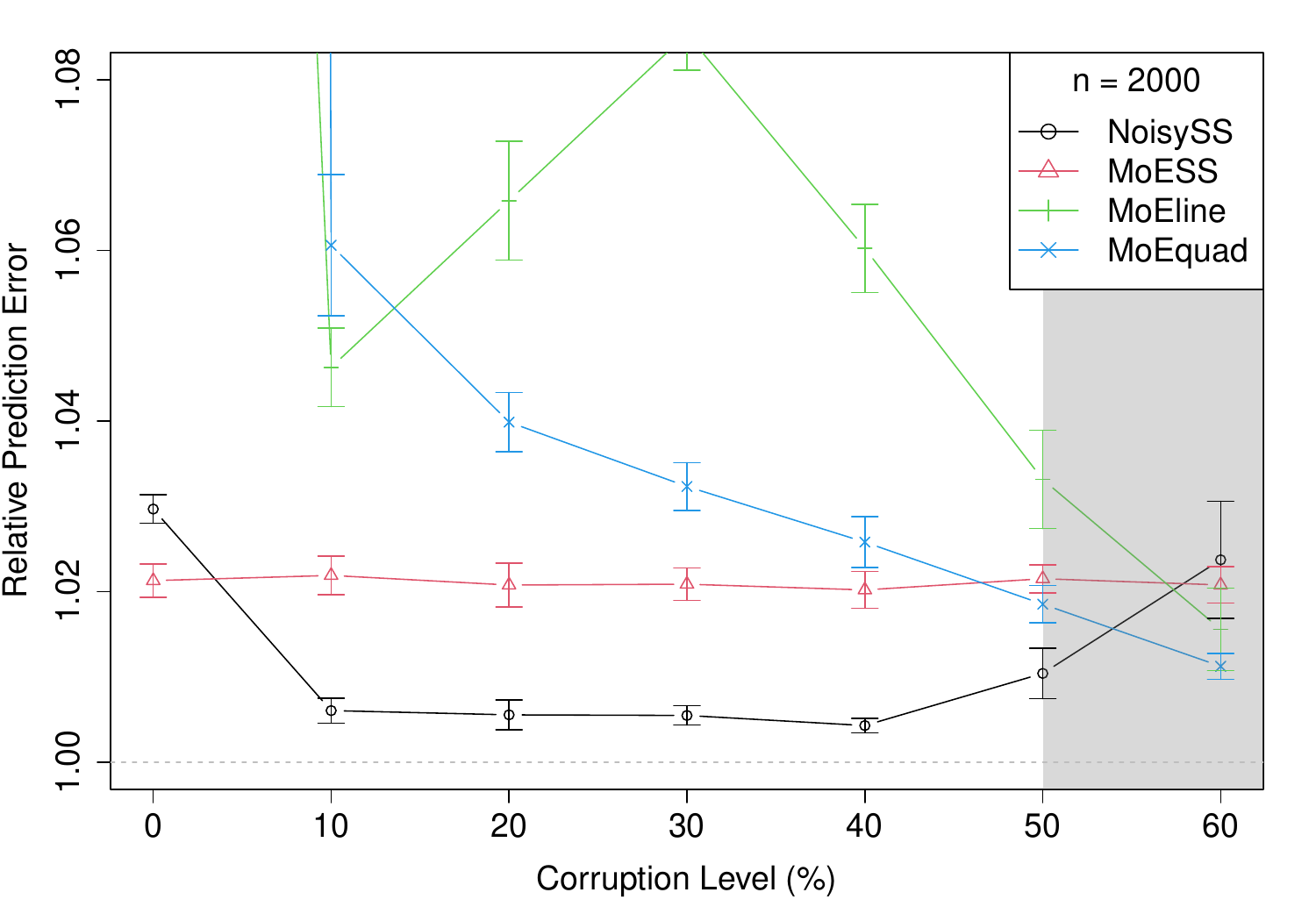}
        \caption{Average of RPE Comparison}
        \label{fig:fixn_y}
    \end{subfigure}
    \caption{Plots of average of MSE and RPE. The gray shaded region represents where we expect our proposed method, NoisySS, to break down. (Error bars represent  95\% confidence intervals.)}
    \label{fig:fixn}
\end{figure}

We repeat the above procedure $50$ times, and report the averages of each measure across these $50$ repetitions in Table \ref{tab:fixn_beta} and Table \ref{tab:fixn_y}. In addition to the tables, we provide plots in Figure \ref{fig:fixn} with error bars representing a 95\% confidence interval. These plots offer a visual representation of the tables, allowing for easier comparison of the estimators' performances.

\begin{table}[h!t]
      \spacingset{1}
\centering
\begin{tabular}{lrrrr}
\toprule
Noise level & \multicolumn{1}{c}{NoisySS} & \multicolumn{1}{c}{MoESS} & \multicolumn{1}{c}{MoEline} & \multicolumn{1}{c}{MoEquad} \\
\midrule
{0\%} & 0.014 (0.001) & 0.012 (0.002) & 1.232 (0.088) & 2.005 (0.175) \\
{10\%} & 0.012 (0.001) & 5.887 (0.547) & 4.379 (0.309) & 6.817 (1.231) \\
{20\%} & 0.013 (0.001) & 10.332 (1.204) & 2.869 (0.103) & 3.847 (0.955) \\
{30\%} & 0.013 (0.001) & 16.151 (1.738) & 2.665 (0.103) & 2.349 (0.387) \\
{40\%} & 0.020 (0.002) & 20.897 (2.091) & 2.227 (0.123) & 1.502 (0.112) \\
{50\%} & 0.718 (0.184) & 20.776 (1.597) & 1.600 (0.092) & 1.158 (0.123) \\
{60\%} & 8.417 (0.768) & 22.037 (2.108) & 1.269 (0.101) & 1.104 (0.253) \\
\bottomrule
\end{tabular}
\caption{Average of MSE of regression coefficients. The standard error is given in parentheses.}
\label{tab:fixn_beta}
\end{table}

\begin{table}[h!t]
  \spacingset{1}
  \centering

\begin{tabular}{lrrrr}
\toprule
Noise level & \multicolumn{1}{c}{NoisySS} & \multicolumn{1}{c}{MoESS} & \multicolumn{1}{c}{MoEline} & \multicolumn{1}{c}{MoEquad} \\
\midrule
{0\%} & 1.030 (0.001) & 1.021 (0.001) & 1.515 (0.055) & 7.843 (1.599) \\
{10\%} & 1.006 (0.001) & 1.022 (0.001) & 1.046 (0.002) & 1.061 (0.004) \\
{20\%} & 1.006 (0.001) & 1.021 (0.001) & 1.066 (0.004) & 1.040 (0.002) \\
{30\%} & 1.005 (0.001) & 1.021 (0.001) & 1.085 (0.002) & 1.032 (0.001) \\
{40\%} & 1.004 (0.000) & 1.020 (0.001) & 1.060 (0.003) & 1.026 (0.002) \\
{50\%} & 1.010 (0.002) & 1.022 (0.001) & 1.033 (0.003) & 1.019 (0.001) \\
{60\%} & 1.024 (0.003) & 1.021 (0.001) & 1.016 (0.002) & 1.011 (0.001) \\
\bottomrule
\end{tabular}
\caption{Average of RPE. The standard error is given in parentheses.}
\label{tab:fixn_y}
\end{table}

We discuss the simulation results based on the noise level. 
\begin{enumerate}
    \item (Zero noise level) When there is no noise, meaning that $\mathrm P(Z = \tilde Z ) = 1$, MoESS outperforms the others in both measures. In this simulation example, $\hat s(X)$ becomes the Bayes rule as $N=\infty$ and the Bayes error, $\mathrm P(\tilde Z\neq \hat s(X))$, is very low. Therefore, $\hat s(X) = Z$ with very high probability. In this case, the information about $Z$ can be almost perfectly recovered from $\hat s(X)$. Since MoESS implicitly assumes that $\hat s(X) = Z$ and utilizes $\hat s$ information, it performs better than the others. Following MoESS, our proposed method NoisySS shows good performance. NoisySS utilizes $\hat s$ but assumes that $\hat s$ and $Z$ could be different. In contrast, the supervised MoE methods do not utilize $\hat s$ information. 
    \item (Moderate noise level) At the corruption level of 10\% to 50\%, NoisySS performs well compared to the others in both measures. When comparing the MSE of the regression coefficients, the MSE for MoESS worsens, while the MSE for NoisySS remains relatively steady. This is because our model expects $\tilde Z$ to be noisy, so it only utilizes a subset of samples for estimating the experts. In contrast, MoESS targets the wrong parameters, as discussed in Section~\ref{subsec:lit_method}. When comparing RPE, both NoisySS and MoeSS' RPE remain steady as the corruption level increases. MoESS maintains steady RPE because it targets the correct mean, even with some noise. However, MoESS performs worse than NoisySS because it does not estimate the correct experts, $\mathrm P(Y|X, Z=k)$, but instead estimates $\textstyle\sum_{k=1}^K \pi_{k|\tilde k} \mathrm P(Y|X, Z=k)$, which has a larger variance due to the mixing of distributions.     
    \item (High noise level) The supervised MoE methods, MoEline and MoEquad, perform well when the noise level is high. If the noise level exceeds 50\%, it indicates that the information on $\tilde Z$ is not helpful in identifying $Z$. Thus, we do not expect NoisySS to do well in this setting.  Since the supervised MoE methods do not transfer any side information and rely only on labeled data, they perform better than the other methods under these conditions. The reason MoEquad performs better than MoEline is related to sample size, which is discussed in the next subsection.
\end{enumerate}

\subsection{Varying label data sizes}

We generate training data with varying sample sizes of 300, 500, 1,000, and 2,000, and test data with a sample size of 20,000. We fix the corruption level at 20\%. For each sample size, we compute each estimator's MSE and RPE, repeating the process 50 times. We report the average MSE and RPE over the 50 replications along with the standard error.

\begin{figure}[h!t]
    \centering
    \begin{subfigure}[b]{0.48\textwidth}
        \centering
        \includegraphics[width=\textwidth]{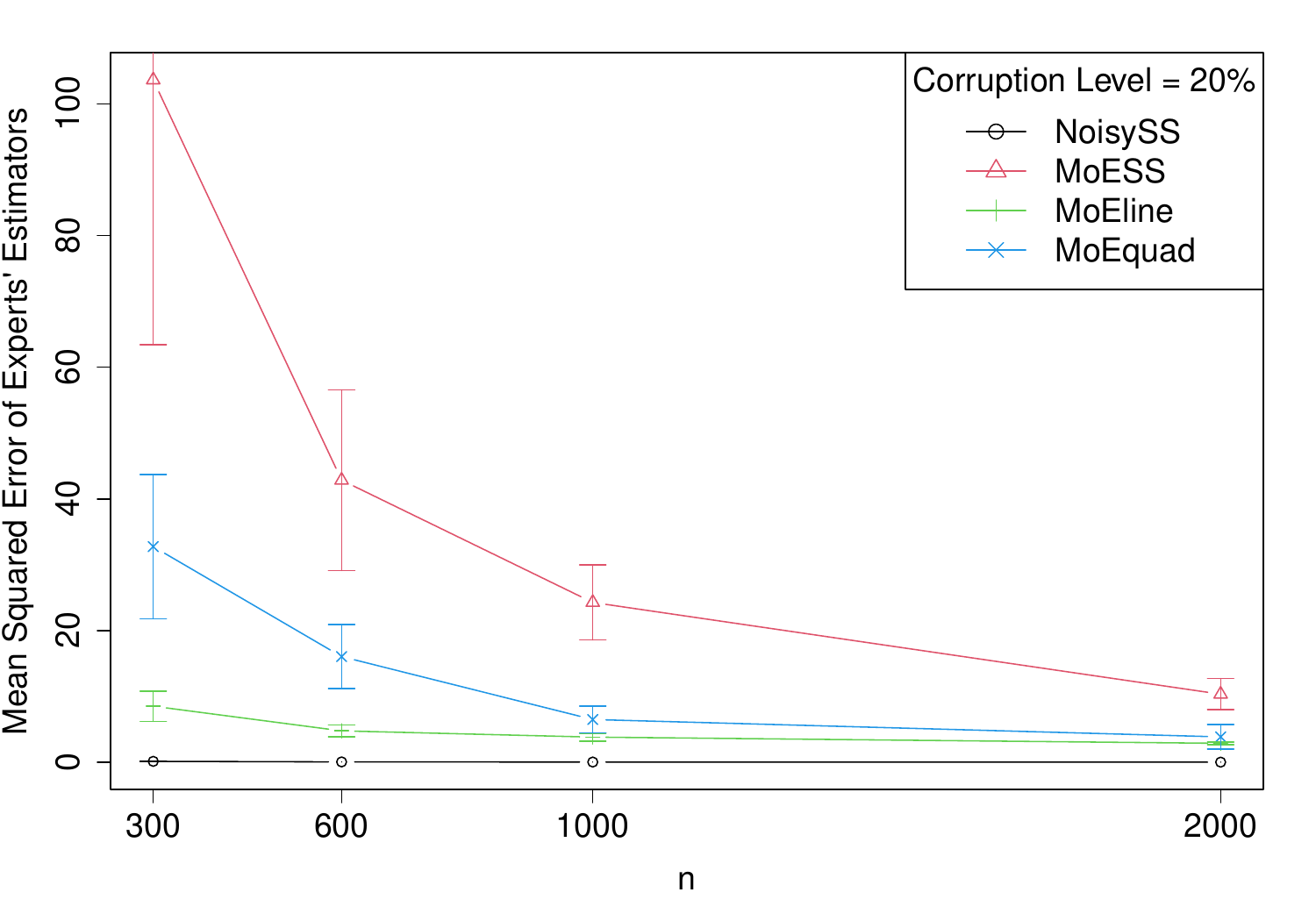}
        \caption{Average of MSE Comparison}
        \label{fig:fixp_beta}
    \end{subfigure}
    \hfill
    \begin{subfigure}[b]{0.48\textwidth}
        \centering
        \includegraphics[width=\textwidth]{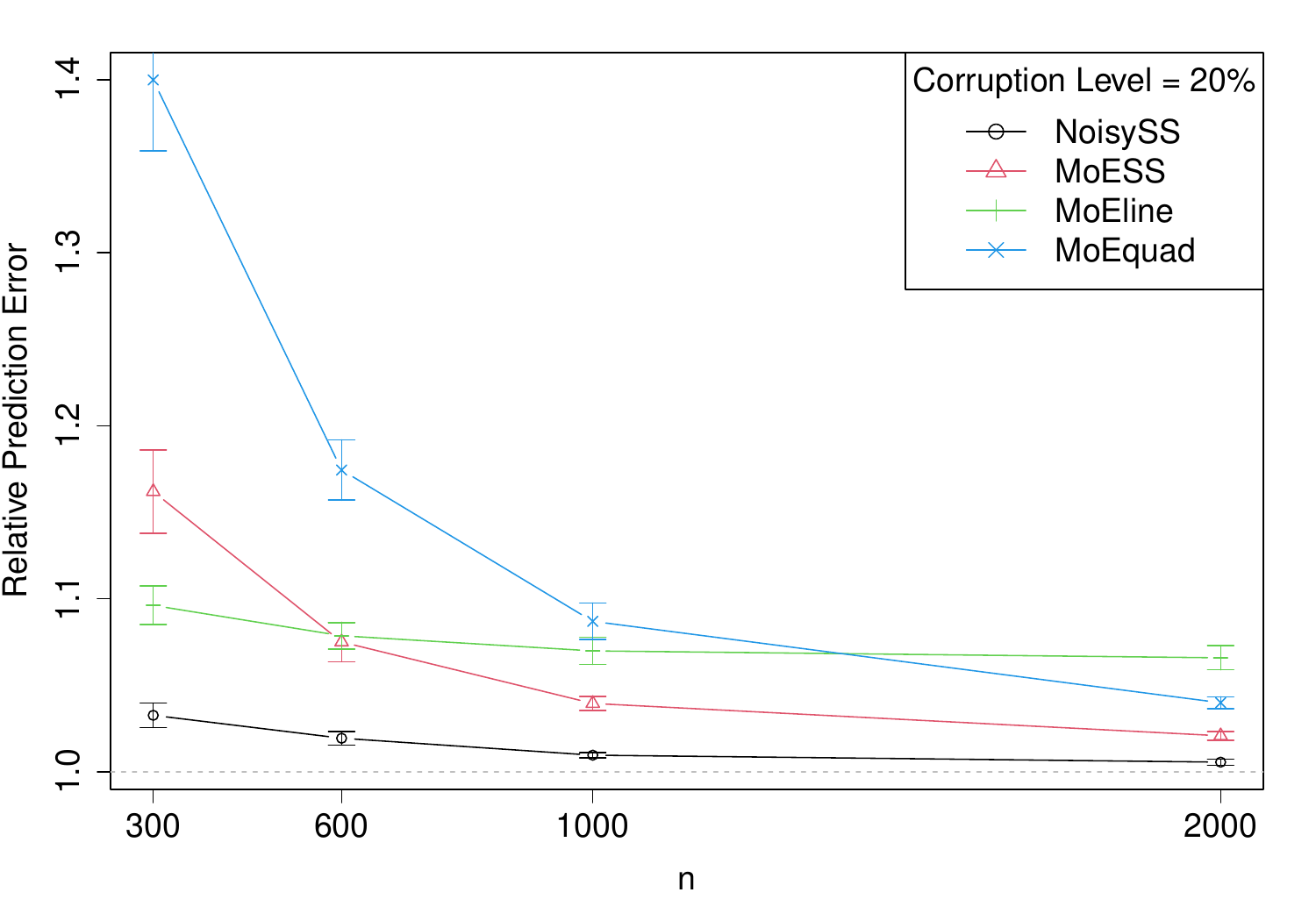}
        \caption{Average of RPE Comparison}
        \label{fig:fixp_y}
    \end{subfigure}
    \caption{Plots of average of MSE and RPE. (Error bars represent  95\% confidence intervals.)}
    \label{fig:fixp}  
\end{figure}

The results are presented in Table \ref{tab:fixp_beta} and Table \ref{tab:fixp_y}. Additionally, Figure \ref{fig:fixp} provides plots with error bars representing a 95\% confidence interval. 

\begin{table}[h!t]
\spacingset{1}
\centering
\begin{tabular}{lrrrr}
\toprule
$n$ & NoisySS & MoESS & MoEline & MoEquad \\
\midrule
{300}  & 0.131 (0.012) & 103.626 (20.527) & 8.510 (1.174) & 32.746 (5.585) \\
{600}  & 0.048 (0.004) & 42.837 (7.002) & 4.763 (0.463) & 16.042 (2.480) \\
{1,000} & 0.028 (0.002) & 24.266 (2.899) & 3.816 (0.304) & 6.486 (1.047) \\
{2,000} & 0.013 (0.001) & 10.332 (1.204) & 2.869 (0.103) & 3.847 (0.955) \\
\bottomrule
\end{tabular}
\caption{Average of MSE. The standard error is given in parentheses.}
\label{tab:fixp_beta}
\end{table}

\begin{table}[h!t]
\spacingset{1}
\centering
\begin{tabular}{lrrrr}
\toprule
$n$ & NoisySS & MoESS & MoEline & MoEquad \\
\midrule
{300}  & 1.033 (0.004) & 1.162 (0.012) & 1.096 (0.006) & 1.400 (0.021) \\
{600}  & 1.019 (0.002) & 1.075 (0.006) & 1.079 (0.004) & 1.174 (0.009) \\
{1,000} & 1.010 (0.001) & 1.039 (0.002) & 1.070 (0.004) & 1.087 (0.005) \\
{2,000} & 1.006 (0.001) & 1.021 (0.001) & 1.066 (0.004) & 1.040 (0.002) \\
\bottomrule
\end{tabular}
\caption{Average of RPE. The standard error is given in parentheses.}
\label{tab:fixp_y}
\end{table}

We have three observations from the simulation results. 
\begin{enumerate}
    \item NoisySS performs well across different sample sizes compared to others. NoisySS's performance is particularly better when the sample size is small. However, the relative improvement decreases as the sample size increases.
    \item For all methods, as the sample size increases, both the estimation of experts and prediction performance improve.
    \item Comparing the supervised MoE methods, MoEquad gives better performance when there is a larger sample size. This may be because MoEquad require more parameters to estimate, so a sufficient sample size is necessary.
\end{enumerate}

\section{Data Applications}\label{sec:real}
We present two data examples to demonstrate our approach. In Section~\ref{sec:real.bank}, the banknote dataset is used primarily to illustrate the proposed method, while Section~\ref{sec:real.star} employs stellar data to compare prediction performance between the methodologies. {\color{black}For these analyses, we assume that the error terms follow a normal distribution.}

\subsection{Banknote Data}\label{sec:real.bank}
The banknote data contain 100 genuine and 100 counterfeit Swiss banknotes, as described in \cite{flury1988multivariate} and available in the \texttt{MASS} package in R. The data include measurements of six different locations on the banknote. While this data set is commonly used for classification tasks (distinguishing between genuine and counterfeit notes), we consider a different task. We predict the length of the diagonal of a banknote ($Y$) using the length ($X_1$) and the distance from the bottom to the inner frame ($X_2$). 
We choose these variables as response and predictor variables because this task clearly illustrates our method compared to MoESS. Thus, this analysis is solely for illustration purposes. Conceptually, we expect $Z$ to represent the banknote's status (whether it is genuine or not) and $\tilde Z$ to represent the labels of clusters defined by the banknote size information.

This prediction is performed when $X_1$ and $X_2$ values are present for all banknotes ($N=200$), but a few values of $Y$ are missing ($n<N$). Additionally, the banknote's status information (whether a banknote is genuine or counterfeit) is taken for our purposes to be unknown. The goal is to predict the missing diagonal values of the banknotes.

\begin{figure}[h!t!]
    \centering
    \begin{subfigure}[b]{0.32\textwidth}
        \centering
        \includegraphics[width=\textwidth]{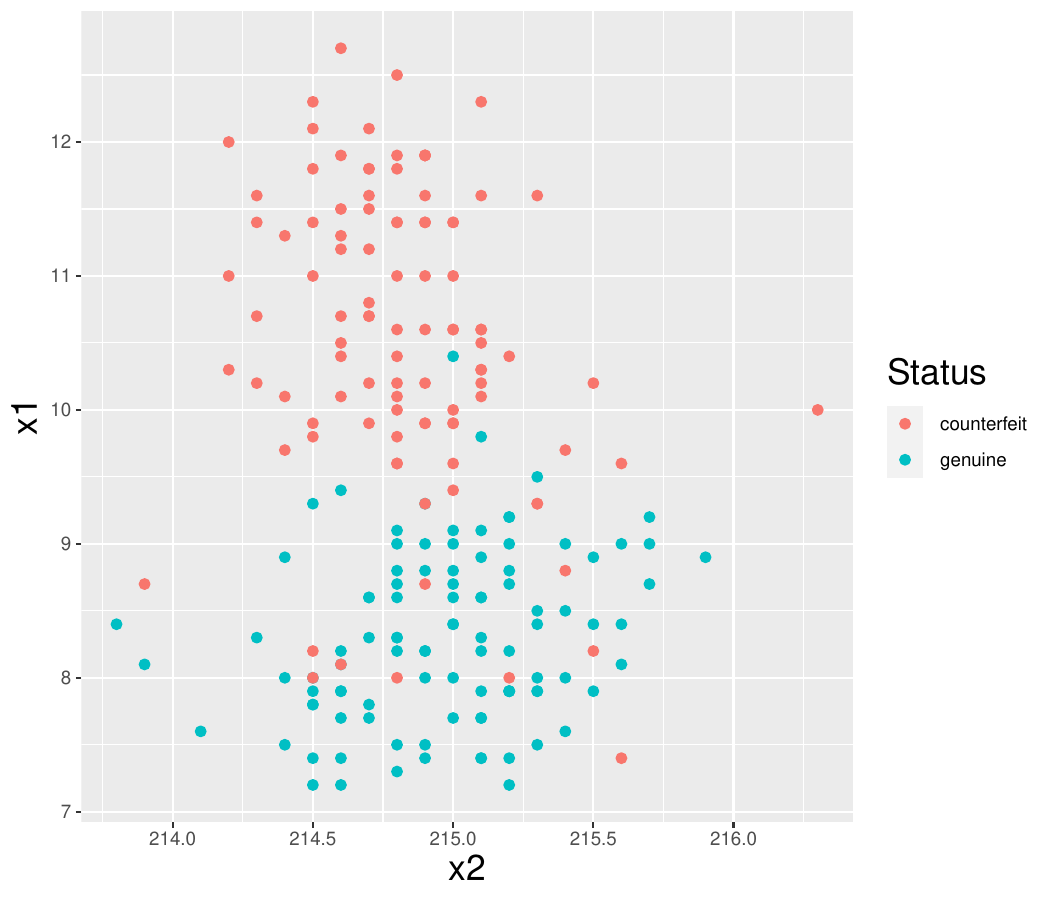}
        \caption{Scatter plot of $X_1$ on $X_2$}
        \label{fig:banknote_true}
    \end{subfigure}
    \hfill
    \begin{subfigure}[b]{0.32\textwidth}
        \centering
        \includegraphics[width=\textwidth]{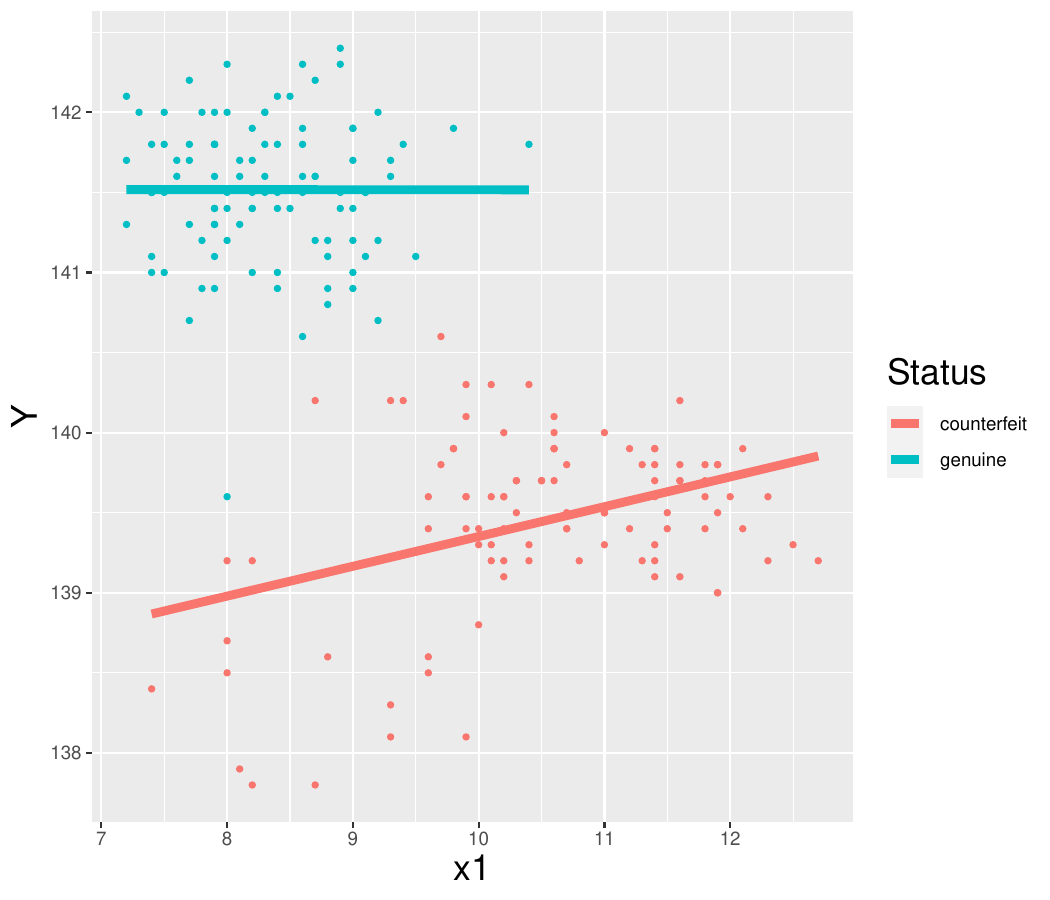}
        \caption{Scatter plot of $Y$ on $X_1$}
        \label{fig:banknote_true_reg1}
    \end{subfigure}
    \hfill
    \begin{subfigure}[b]{0.32\textwidth}
        \centering
        \includegraphics[width=\textwidth]{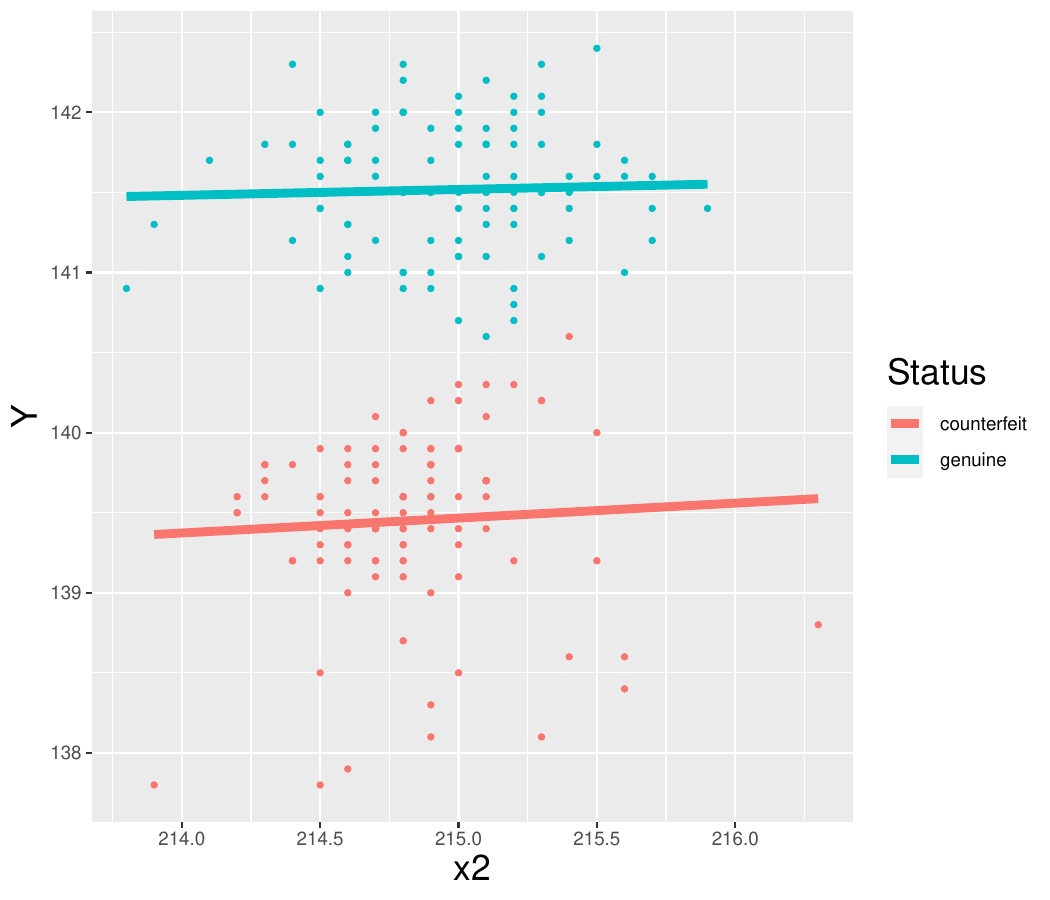}
        \caption{Scatter plot of $Y$ on $X_2$}
        \label{fig:banknote_true_reg2}
    \end{subfigure}
    \caption{Pairwise scatter plots for ($Y,X_1, X_2$) with points colored by $Z$}
    \captionsetup{justification=centering} 
    \vspace{-0.3cm}
    \caption*{Note: The lines represent the linear regression lines for each class.} 
\end{figure}

Based on Figures \ref{fig:banknote_true_reg1} and \ref{fig:banknote_true_reg2}, which contain scatter plots of $Y$ on $X_2$ and $X_1$, respectively, for all 200 banknotes, fitting separate linear regression models for $Y$ on the two predictors might be a good practice if the status information were known. 

\begin{figure}[h!t!]
    \centering
    \begin{subfigure}[b]{0.32\textwidth}
        \centering
        \includegraphics[width=\textwidth]{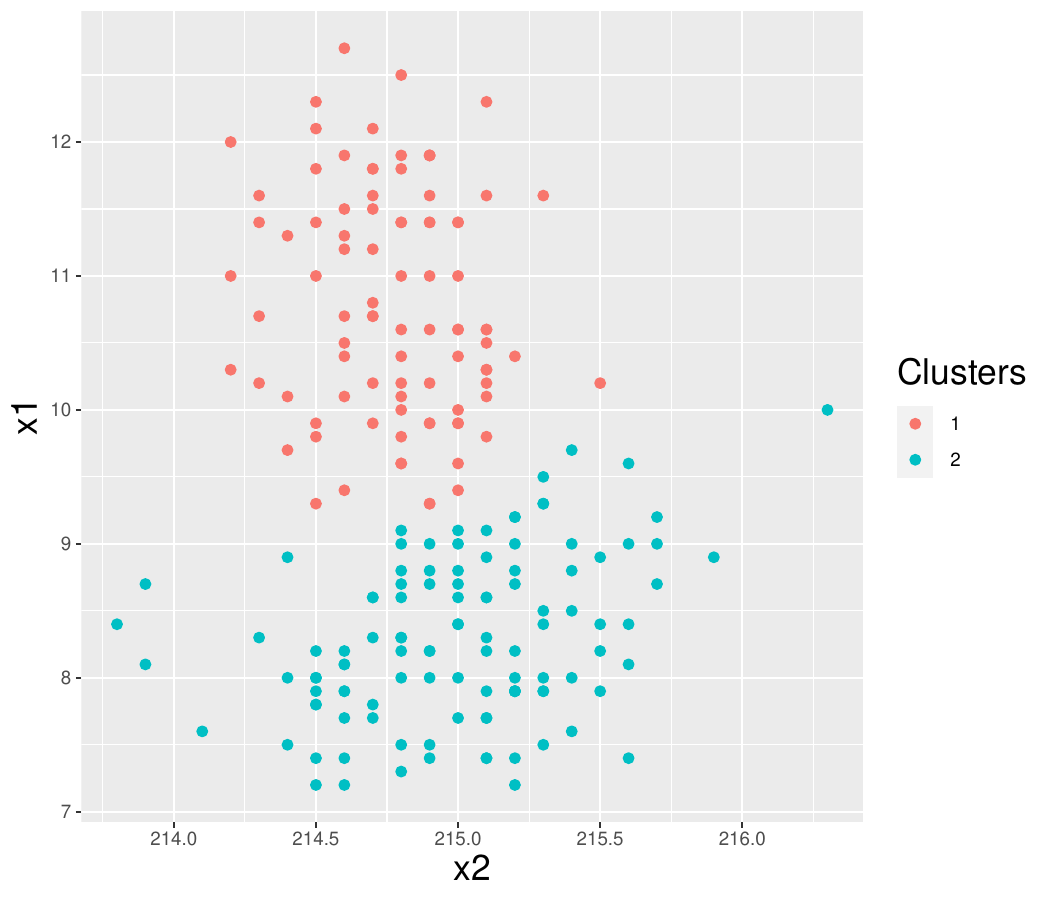}
        \caption{Scatter plot of $X_1$ on $X_2$}
        \label{fig:x_gmm}
    \end{subfigure}
    \hfill
    \begin{subfigure}[b]{0.32\textwidth}
        \centering
        \includegraphics[width=\textwidth]{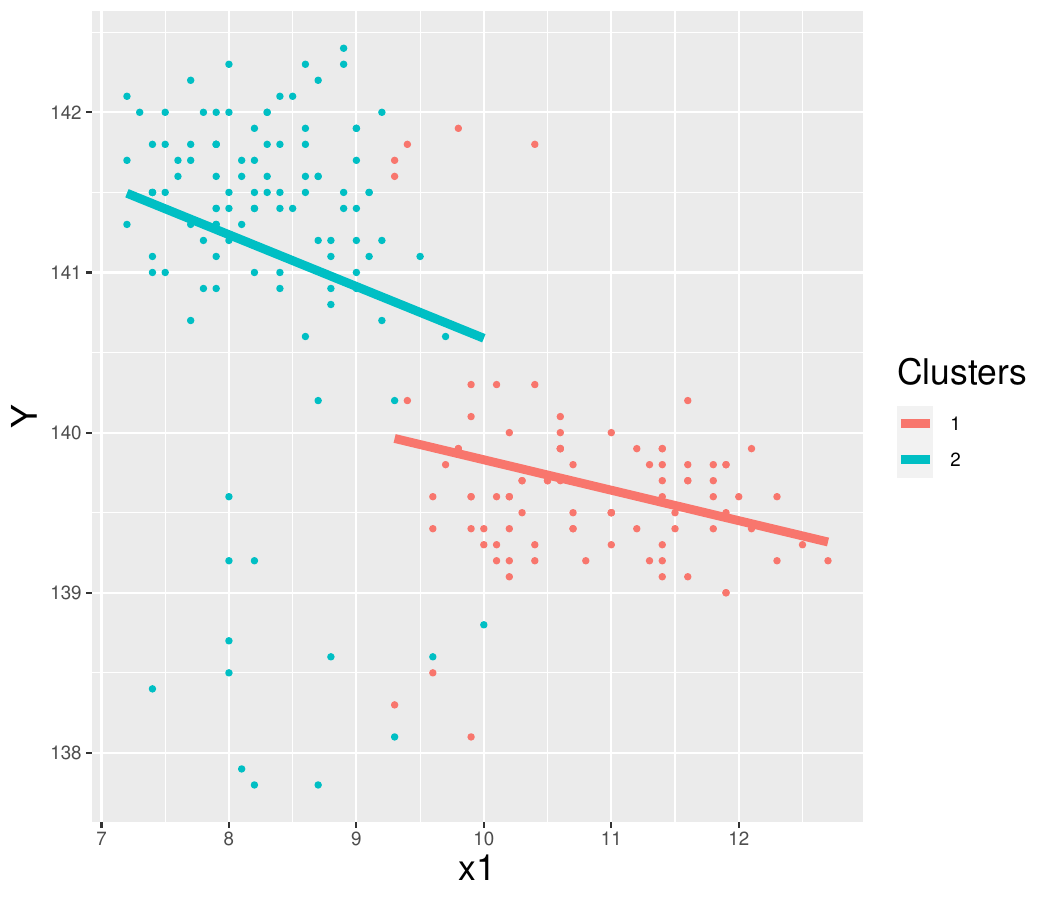}
        \caption{Scatter plot of $Y$ on $X_1$}
        \label{fig:yx1_gmm}
    \end{subfigure}
    \hfill
    \begin{subfigure}[b]{0.32\textwidth}
        \centering
        \includegraphics[width=\textwidth]{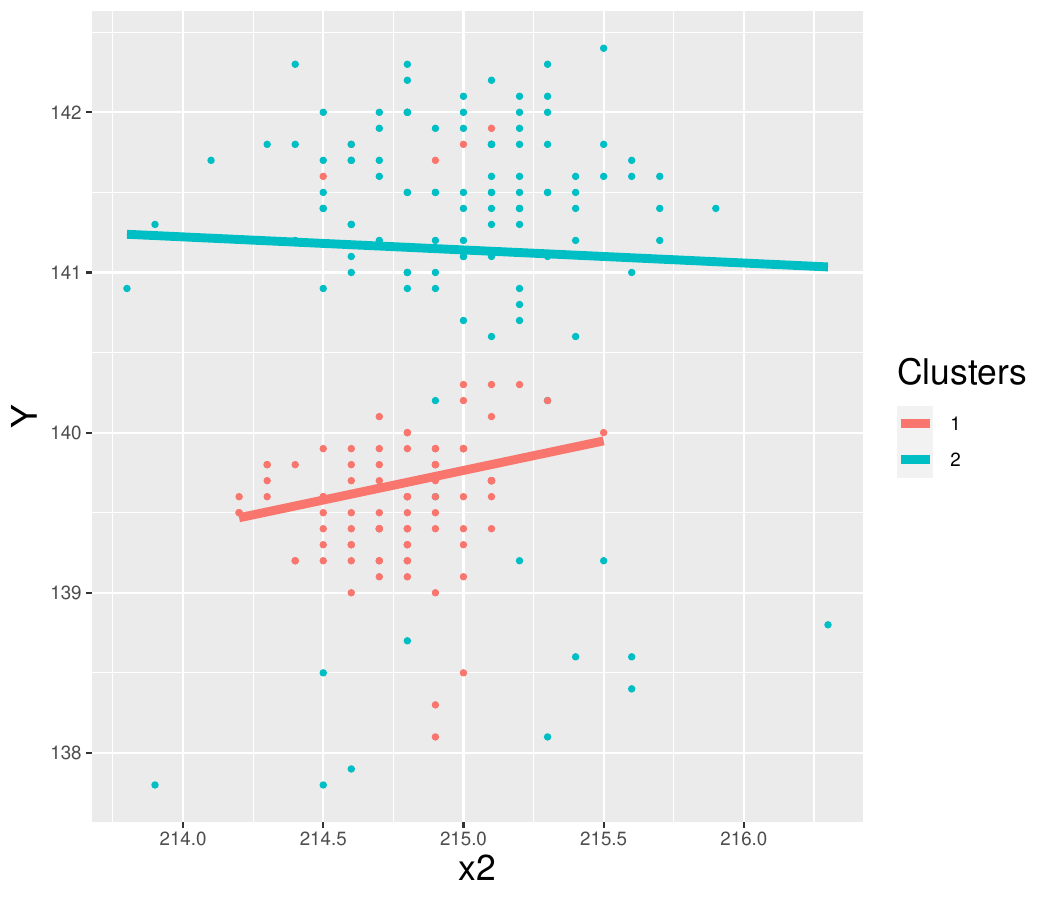}
        \caption{Scatter plot of $Y$ on $X_1$}
        \label{fig:yx2_gmm}
    \end{subfigure}
        \caption{Pairwise scatter plots for ($Y,X_1, X_2$) with points colored by $\hat s$}
    \label{fig:banknote_gmm}
        \captionsetup{justification=centering} 
    \vspace{-0.3cm}
    \caption*{Note: The lines represent the linear regression lines for each class.} 
\end{figure}

However, in our scenario, the class information is unknown. In MoESS, the class information is imputed on the space of $X$ using a Gaussian mixture model, as shown in Figure \ref{fig:x_gmm}. Then, for each imputed class, linear regressions are fitted. However, as we notice from Figures \ref{fig:yx1_gmm} and \ref{fig:yx2_gmm}, there are a few observations that are wrongly classified, which may lead to inaccuracies in the regression models. In contrast, our proposed method uses robust linear regression for estimating the regression line, which eliminates observations that are outliers or potentially wrongly classified. Thus, we expect to improve the accuracy of the model.

Using banknote data, we compare the prediction performance of four estimators: NoisySS, MoESS, MoEline, and MoEquad. We set $K=2$. 

\begin{figure}[h!t!]
    \centering
    \begin{subfigure}[b]{0.8\textwidth}
        \centering
        \includegraphics[width=\textwidth]{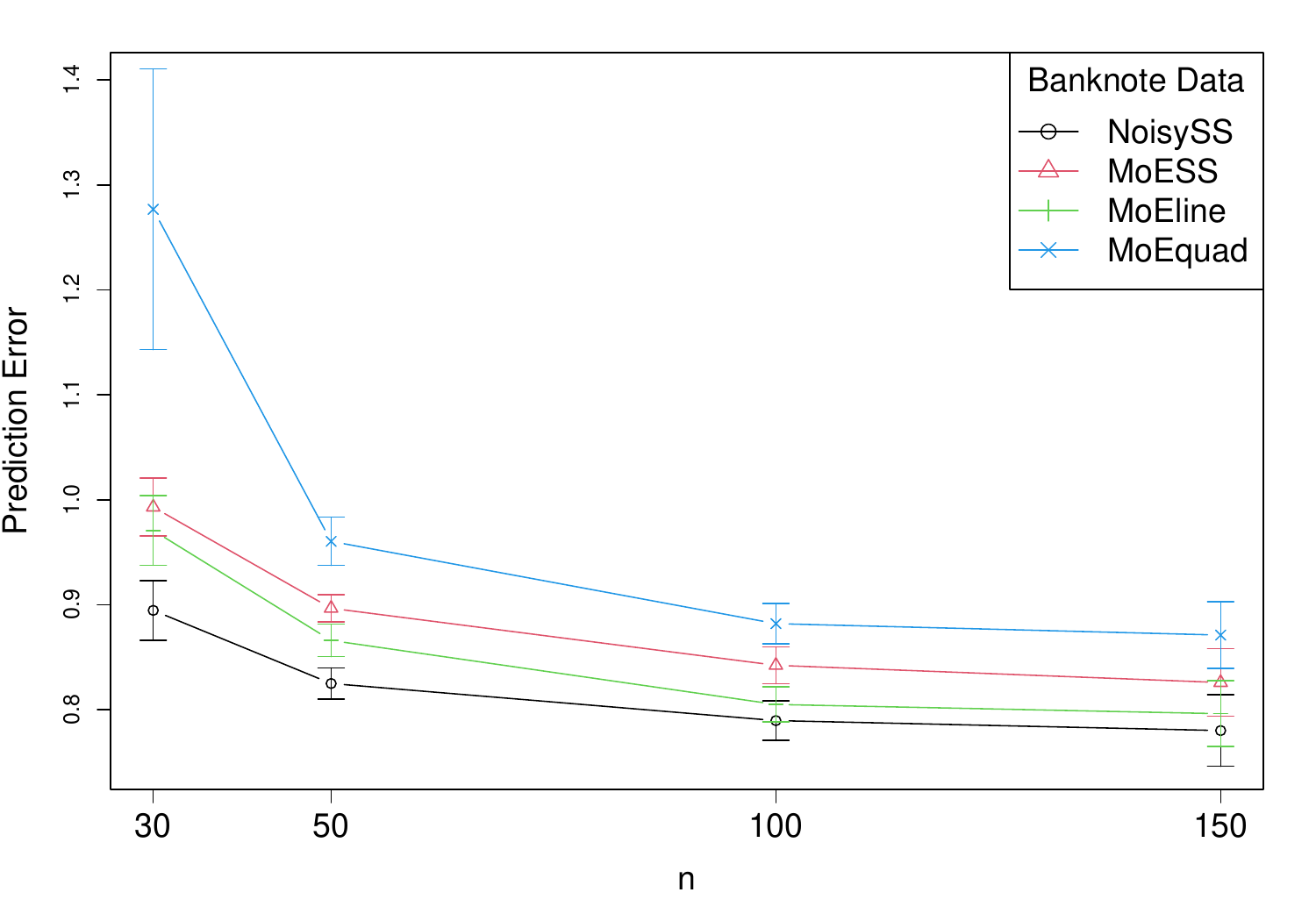}
    \end{subfigure}
    \caption{Average PE on banknote data}
    \label{fig:banknote_PE}
\end{figure}

We randomly select $n$ banknotes for the training data and predict the $Y$ values for the remaining $200-n$ banknotes. As the true expected value is unknown, we calculate the prediction error on the test data:
$$\text{PE}(\hat y^{\text{test}}) = \textstyle\sum_{i=1}^{200-n}  ( y_i^{\text{test}} - \hat y_i^{\text{test}} ) ^2 / (200-n).$$ We repeat the process 200 times and report the average PE with the standard error, which is summarized in Table \ref{tab:banknote_PE} and Figure \ref{fig:banknote_PE} with varying $n$ of 30, 50, 100, and 150. 

\begin{table}[h!t!]
\spacingset{1}
\centering
\begin{tabular}{lrrrr}
\toprule
$n$ & {NoisySS} & {MoESS} & {MoEline} & {MoEquad} \\
\midrule
30  & 0.895 (0.014) & 0.993 (0.014) & 0.971 (0.017) & 1.277 (0.068) \\
50  & 0.825 (0.008) & 0.897 (0.007) & 0.866 (0.008) & 0.960 (0.012) \\
100 & 0.790 (0.010) & 0.842 (0.009) & 0.805 (0.009) & 0.882 (0.010) \\
150 & 0.780 (0.017) & 0.826 (0.016) & 0.796 (0.016) & 0.871 (0.016) \\
\bottomrule
\end{tabular}
\caption{Mean and Standard Error of Prediction Error for Different Estimators and Sample Sizes}
\label{tab:banknote_PE}
\end{table}

When the labeled data size $n$ is small, NoisySS's prediction performance is better than other methods. However, as $n$ increases, the relative improvement decreases. This observation aligns with the simulation results. When $n \geq 100$, the relative improvement is not significant. This may be because the amount of transferable information from unlabeled data decreases as the labeled data size ($n$) becomes large relative to the unlabeled data size ($N$). In other words, in this example, as the unlabeled data size is small, when $n$ gets larger, it starts to resemble a supervised learning problem rather than the traditional semi-supervised learning problem, which assumes $N \gg n$.

\begin{remark}\label{rmk:assump}
Recall from Section \ref{sec:model} our  modeling assumption that $X \ind Z | \tilde Z$.  We can probe this assumption in this example. Since $\tilde Z$ is unknown, we consider $\hat s$ as an approximation for $\tilde Z$. Given the cluster $\hat s = 2$, represented by the black points in Figure \ref{fig:x_gmm}, we do not observe any pattern in the transitions from $\hat s$ to $Z$ that is explained by $X$ in Figure \ref{fig:banknote_true}, implying that the conditional independence assumption appears to be satisfied. 
However, given that the cluster $\hat s = 1$ has so few points transitioning from one to the other, it is difficult to determine whether the conditional independence assumption is violated.
\end{remark}

\subsection{Stellar Data}\label{sec:real.star}
In this subsection, we use data from the Sloan Digital Sky Survey (SDSS) Stellar Parameter Pipeline (SSPP) \citep{astroMLText}, which can be downloaded using the \texttt{fetch\_sdss\_sspp} function from the \texttt{astroML} library in Python, available at \url{https://github.com/astroML/astroML}. These data contain various stellar parameters for stars observed by the SDSS, which are often used for astronomical analyses. We use cleaned SDSS SSPP data, where objects with extreme values are removed, available via \texttt{fetch\_sdss\_sspp(cleaned=TRUE)}. The detailed cleaning process is described in their data documentation. 
The cleaned data consist of $N=76,127$ observations. 

The iron-to-hydrogen ratio ($Y$) is a crucial indicator of a star's metallicity, helping us understand the star's formation environment and age. Metallicity is interrelated with the alpha-to-iron ratio ($X_1$) and effective temperature ($X_2$), and these relationships differ for various types of stars. In this example, $\tilde Z$ can be thought of as the different types of stars. 

We use the cleaned SDSS SSPP data to predict $Y$ using $X_1$ and $X_2$. Each variable is standardized to have mean of 0 and standard deviation of 1. 

We compare the prediction performance of the same four estimators as in the previous subsection. For NoisySS and MoESS, when we fit a Gaussian mixture model on the predictor variables of the unlabeled data, we determine the number of components by examining the scree plot of the number of clusters versus the Bayesian Information Criterion (BIC). We choose the elbow point on the plot and select $K=10$. 

\begin{figure}[h!t!]
    \centering
    \begin{subfigure}[b]{0.8\textwidth}
        \centering
        \includegraphics[width=\textwidth]{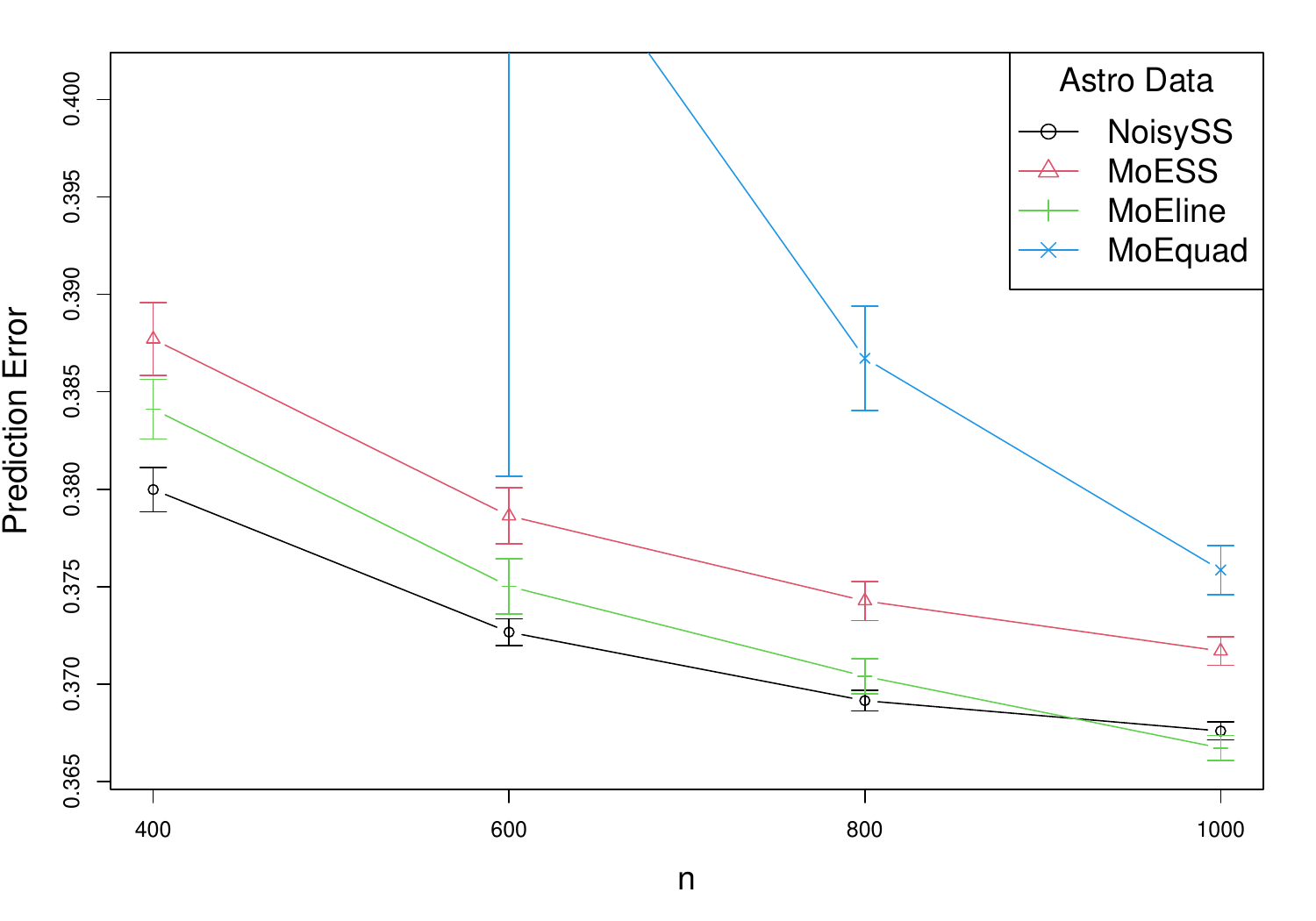}
    \end{subfigure}
    \caption{Average PE on stellar data}
    \label{fig:stellar_PE}
\end{figure}

We randomly select $n$ stars for the training data and predict the $Y$ values for the remaining $N-n$ stars and calculate the prediction error on the test data. We repeat the process 200 times and report the average PE with the standard error, which is summarized in Table \ref{tab:stellar_PE} and Figure \ref{fig:stellar_PE} with varying $n$ of 400, 600, 800, and 1,000.

\begin{table}[h!t!]
\spacingset{1}
\centering
\begin{tabular}{lrrrr}
\toprule
$n$ & {NoisySS} & {MoESS} & {MoEline} & {MoEquad} \\
\midrule
400  & 0.380 (0.001) & 0.388 (0.001) & 0.384 (0.001) & 0.423 (0.002) \\
600  & 0.373 (0.000) & 0.379 (0.001) & 0.375 (0.001) & 0.412 (0.016) \\
800  & 0.369 (0.000) & 0.374 (0.001) & 0.370 (0.000) & 0.387 (0.001) \\
1000 & 0.368 (0.000) & 0.372 (0.000) & 0.367 (0.000) & 0.376 (0.001) \\
\bottomrule
\end{tabular}
\caption{Mean and Standard Error of Prediction Error for Different Estimators and Sample Sizes}
\label{tab:stellar_PE}
\end{table}

When the sample size is small, NoisySS's prediction performance is better than other methods. However, as the sample size grows, the relative improvement diminishes. Again, this result aligns with our simulation results. When the labeled sample size is large, MoEline performs slightly better than NoisySS, suggesting that the connection between $Z$ and $\tilde Z$ may be rather noisy.

\section{Discussion}\label{sec:dis}
In this paper, we introduce the noisy mixture of experts model, which allows for flexibility by permitting the clusters on $X$ to differ from those on $Y|X$. We then develop a semi-supervised approach for this model. Our approach leverages a large amount of unlabeled data to assign labeled data to clusters based on $X$. To address the discrepancies between the clusters on $X$ and those on $Y|X$, we employ a least trimmed squares estimation-based algorithm to trim misaligned data on the labeled data. In our theoretical analysis, we characterize the conditions under which the proposed approach is effective.

A potential direction for future research is to relax the conditional independence assumption, allowing the transition probability from $\tilde Z$ to $Z$ to depend on $X$. Without this assumption, additional technical challenges arise. For instance, if we model the transition probability using a standard multinomial logistic regression, we would need to estimate $K^2\cdot (p+1)$ parameters. Addressing the issues caused by this large number of parameters might require incorporating a penalty term.

Another direction for further research is to improve the accuracy of the estimation process by utilizing trimmed samples. In this paper, the $k$-th expert is trained using approximately $\alpha n\mathrm P(\tilde Z=k )$ samples from the labeled data. Although the remaining samples may not provide sufficient information about the parameters of the $k$-th expert, they still hold relevant information if the gate probability is non-zero. While the least trimmed squares algorithm discards these samples, finding ways to leverage their information could improve estimation accuracy.

\section*{Acknowledgments}

This work was supported by a grant by the Simons Collaboration on Computational Biogeochemical Modeling of Marine Ecosystems/CBIOMES (Grant ID: 549939 to JB). The authors acknowledge the Center for Advanced Research Computing (CARC) at the
University of Southern California for providing computing resources that have contributed to the research
results reported within this publication. URL: \url{https://carc.usc.edu}. We thank Elizabeth C. Atwood, Bror J{\"o}nsson, and Thomas Jackson for many useful conversations in understanding chlorophyll prediction models used in oceanography and Matias Salibian Barrera for a helpful email about robust estimators.

\bibliographystyle{asa.bst} 
\bibliography{ref}

\end{document}